\documentclass[journal, comsoc]{IEEEtran}
\IEEEoverridecommandlockouts
\usepackage{amsmath,amssymb,amsfonts,amsthm}
\usepackage{pgfplots}
\usepackage{pgfplotstable}
\pgfplotsset{compat=1.18}
\usepackage[hidelinks]{hyperref}
\usepackage{orcidlink}
\usepackage{physics}
\usepackage{mleftright}
\usepackage{mathtools}
\usepackage{tikz}
\usetikzlibrary{arrows.meta}
\usetikzlibrary{matrix}
\usepgflibrary{arrows}
\usetikzlibrary{plotmarks}
\usepackage{stmaryrd}
\usepackage[Tol, keep-defaults]{colorblind}
\usepackage[T1]{fontenc}
\usepackage{siunitx}
\usepackage{microtype}
\usepackage[caption=false,font=footnotesize]{subfig}

\def\BibTeX{{\rm B\kern-.05em{\sc i\kern-.025em b}\kern-.08em
    T\kern-.1667em\lower.7ex\hbox{E}\kern-.125emX}}

\newcommand{\ticklabelsize}[0]{\footnotesize}

\newcommand{\EsToEb}[2]{{\thisrow{#1} - 10*ln(#2)/ln(10)}}

\newcommand{\figref}[1]{\figurename~\ref{#1}}

\makeatletter
\newcommand{\fixed@sra}{$\vrule height 2\fontdimen22\textfont2 width 0pt\shortrightarrow$}
\newcommand{\shortarrow}[1]{%
  \mathrel{\text{\rotatebox[origin=c]{\numexpr#1*45}{\fixed@sra}}}
}
\makeatother

\pgfplotsset{
    rrsls/.style={solid},
    rrhls/.style={densely dashed},
    dirls/.style={densely dotted},
    gg2ls/.style={dashdotted},
    rrs colour/.style={color=T-Q-HC3},
    rrh colour/.style={color=T-Q-HC2},
    dir colour/.style={color=T-Q-HC4},
    xi 0.01/.style={
        color=T-Q-V2,
    },
    xi 0.03/.style={
        color=T-Q-V4
    },
    rrh mark/.style={mark options={solid}, mark=square, mark repeat=2},
    rrs mark/.style={mark options={solid}, mark=o, mark phase=2, mark repeat=2},
    dir mark/.style={mark options={solid}, mark=triangle, mark repeat=2},
    rrh mark*/.style={mark options={solid}, mark=square*, mark repeat=2},
    rrs mark*/.style={mark options={solid}, mark=*, mark phase=2, mark repeat=2},
    dir mark*/.style={mark options={solid}, mark=triangle*, mark repeat=2},
}
\title{Soft-Decoding Reverse Reconciliation\\in Discrete-Modulation CV-QKD
    \thanks{This work has been partially supported by PNRR MUR project PE0000023-NQSTI, by the QUASAR project and by the SMARTQKD scholarship funded by SMA-RTY Italia SRL and CNR-IEIIT.}
    \thanks{This paper is an extended version of our previous work presented at the 14th International ITG Conference on Systems, Communications and Coding (SCC), Karlsruhe, Germany, 2025 \cite{origlia2025soft}.}
}

\author{
    \IEEEauthorblockN{Marco Origlia%
        \orcidlink{0000-0003-4901-4350}%
        , Emanuele Parente%
        \orcidlink{0009-0009-7981-1052}%
        , and Marco Secondini%
        \orcidlink{0000-0001-5883-5850}, \emph{Senior Member, IEEE}%
  }
}

\newcommand{\matchparen}[1]{\left(#1\right)}
\newcommand{\Abet}[0]{\mathcal{A}}
\newcommand{\Xhat}[0]{\hat{X}}
\newcommand{\xhat}[0]{\hat{x}}

\newcommand{\PP}[1]{\text{P}\{#1\}}
\newcommand{\PPP}[2][]{\text{P}_{#1}(#2)}

\newcommand{\XX}[0]{\mathbf{X}}
\newcommand{\XXhat}[0]{\mathbf{\Xhat}}
\newcommand{\BB}[0]{\mathbf{B}}
\newcommand{\BBhat}[0]{\mathbf{\hat{B}}}

\newcommand{\YY}[0]{\mathbf{Y}}

\newcommand{\vnentr}[1]{\chi\left(#1\right)}

\definecolor{ptb1}{RGB}{187, 187, 187}
\definecolor{ptb2}{RGB}{ 46,  37, 133}
\definecolor{ptb3}{RGB}{ 51, 117,  56}
\definecolor{ptb4}{RGB}{ 93, 168, 153}
\definecolor{ptb5}{RGB}{148, 203, 236}
\definecolor{ptb6}{RGB}{220, 205, 125}

\pgfmathsetmacro{\myN}{.865087}
\pgfmathsetmacro{\myyone}{-2.350028}
\pgfmathsetmacro{\myytwo}{-.350000}
\pgfmathsetmacro{\myythree}{1.649998}
\pgfmathsetmacro{\myyfour}{3.694615}
\pgfmathsetmacro{\myfone}{.008506}
\pgfmathsetmacro{\myftwo}{.343867}
\pgfmathsetmacro{\myfthree}{.606037}
\pgfmathsetmacro{\myffour}{.062901}

\newcommand{\xunitvect}{.825cm}
\newcommand{\yunitvect}{{2*\xunitvect}}

\theoremstyle{remark}
\newtheorem*{remark*}{Remark}
\newtheorem{remark}{Remark}

\usepackage{standalone}
\begin{document}

\maketitle

\begin{abstract}

In continuous-variable quantum key distribution, information reconciliation is required to extract a shared secret key from correlated random variables obtained through the quantum channel. Reverse reconciliation (RR) is generally preferred, since the eavesdropper has less information about Bob's measurements than about Alice's transmitted symbols.
When discrete modulation formats are employed, however, soft information is available only at Bob's side, while Alice has access only to hard information (her transmitted sequence). This forces her to rely on hard-decision decoding to recover Bob's key.

In this work, we introduce a novel RR technique for PAM (and QAM) in which Bob discloses a carefully designed soft metric to help Alice recover Bob's key, while leaking no additional information about the key to an eavesdropper. We assess the performance of the proposed technique in terms of the achievable secret key rate (SKR) and its bounds, showing that the achievable SKR closely approaches the upper bound, with a significant gain over hard-decision RR. Finally, we implement the scheme at the coded level using binary LDPC codes with belief-propagation decoding, assess its bit-error rate through numerical simulations, compare the observed gain with theoretical predictions from the achievable SKR, and discuss the residual gap.
\end{abstract}


\section{Introduction}
Information reconciliation is the process by which two parties, holding correlated random sequences, communicate over a public channel to correct discrepancies between their data and obtain two identical sequences.
This is particularly useful in all applications where these sequences are employed as symmetrical keys,
and the correlated sequences are obtained from data transmission schemes of various natures.
Ensuring secrecy is therefore a fundamental requirement in such kinds of applications.
A possible scenario of shared randomness generation is quantum key distribution (QKD), in which the correlated sequences are obtained from the transmission of quantum states over a quantum channel \cite{Pirandola:20}.

Following the initial phase of random data transmission, reconciliation serves to resolve discrepancies between the correlated sequences.
It is typically assumed that the remote parties are provided with an authenticated noiseless public channel for this purpose.
Finally, a privacy amplification stage is applied to purge any information potentially leaked during the previous phases.

Depending on which party performs error correction, two reconciliation strategies can be distinguished. In direct reconciliation (DR), the sender (Alice) defines the key based on her data and discloses some redundant information, usually a syndrome, associated with it. The receiver (Bob) then uses the shared information, along with his received data, to recover Alice's key.
Conversely, in reverse reconciliation (RR), Bob defines the key based on his received data and discloses the redundant information associated with it, while Alice uses this information and her transmitted data to recover Bob's key.

When a potential attacker is taken into account, the direction of the reconciliation matters.
In fact, the data potentially leaked during transmission usually have larger mutual information with the transmitted data than with the received data.
As a result, RR generally results in lower information leakage compared to DR%
, overcoming the 3dB limitation of DR---which consists of a disruption of the key exchange protocol due to the eavesdropper being able to perfectly reconstruct the secret key by obtaining more than half of the transmitted power---and enabling higher secret key rates (SKR) and longer distances \cite{Pirandola:20,grosshans_reverse_2002}.

For the phase of data transmission, different signaling techniques may be employed.
For instance, in the realm of QKD, the existing protocols may adopt single photons (discrete-variable (DV) QKD)
\cite{bennett_quantum_2014}, or coherent states (continuous-variable (CV) QKD) \cite{grosshans_continuous_2002}.
Unlike DV-QKD, CV-QKD does not require single photon sources and detectors and can use the same devices and modulation/detection schemes commonly employed in
classical coherent optical communications \cite{lodewyck2005controlling}.

The choice of the information carrier is not only a matter of implementation,
as it also affects the kind of variables with which Alice and Bob are provided at the beginning of the reconciliation procedure.
In the GG02 protocol \cite{grosshans_continuous_2002}, Alice and Bob share correlated Gaussian random variables. 
The most well-known reconciliation procedures for Gaussian variables are slice reconciliation \cite{vanassche2004sliced}
and multidimensional reconciliation \cite{leverrier_multidimensional_2008,liu_otfs-based_2025,feng_virtual_2021}.

Despite its theoretical appeal, communication based on Gaussian variables poses several practical challenges,
such as the limited range and finite resolution of modulators
\cite{jouguet2012analysis,djordjevic_discretized_2019}, and the limited reconciliation efficiency \cite{leverrier_unconditional_2009}.
This has motivated the exploration of CV-QKD protocols utilizing discrete constellations \cite{leverrier2009unconditional,hirano2017implementation}.
Among these, a particularly promising approach uses probabilistically-amplitude-shaped quadrature amplitude modulation (PAS-QAM), which allows the protocol to closely approach the SKR achievable by GG02
\cite{notarnicola2023probabilistic,roumestan2024shaped,parente_discrete-modulation_2026}.
When the transmitted symbols are drawn from a discrete constellation, DR can be formulated as a classical error correction problem over an additive white Gaussian noise (AWGN) channel, with slight modifications due to Alice and Bob working within an arbitrary coset of a given code. After Alice publicly reveals the particular coset by disclosing the syndrome of the key (equal to the transmitted sequence in this case), Bob uses this syndrome and the soft information available on his side (the AWGN channel output) to infer the most likely transmitted sequence. For this task, efficient codes and soft-decoding algorithms---such as low-density parity-check (LDPC) codes %
and belief propagation \cite{ryan2009channel, lodewyck_quantum_2007}%
---are available to closely approach channel capacity.
On the other hand, RR---usually preferred over DR, as discussed above---is made more difficult by the lack of soft information on Alice's side. After Bob defines the key based on the received data (e.g., by using thresholds to select the most probable symbols transmitted by Alice) and discloses the corresponding syndrome, Alice has access only to hard information (her transmitted sequence) to recover Bob's decisions through error correction.
In this scenario, where the reconciliation protocol exclusively involves discrete random variables,
secrecy against general attacks is guaranteed by security proofs based on the entropy accumulation theorem \cite{bauml_security_2024,kanitschar_finite-size_2023}. %
However, discarding the soft information contained in Bob's continuous measurement outcomes inevitably reduces the mutual information (MI) between Alice and Bob,
and it is unclear whether a possible reduction of Eve's Holevo information can systematically compensate for this MI loss.
Moreover, \cite{leverrier_information_2023} argues that channel codes closely approaching the capacity of continuous-output channels at very low SNR can be obtained by applying simple constructions, which are not as effective when applied to discrete-output channels. %

For BPSK and QPSK constellations, a soft RR procedure has been proposed by Leverrier \cite{leverrier2009theoretical}, in which Bob discloses the amplitude of the received samples. In this case, the amplitude gives no information to a potential eavesdropper (Eve) about Bob's decisions (which depend only on the phase), but enables Alice to perform soft decoding as in the DR scheme (and with the same efficiency). Unfortunately, this procedure cannot be extended to general discrete constellations (e.g., QAM), for which disclosing the received amplitude would reveal also relevant information about the key.

To overcome this issue, in this work:
\begin{itemize}
\item We introduce a novel RR procedure with soft information (RRS) that can be applied to PAM (and QAM) signaling;
\item We analyze the achievable SKR of the scheme without an eavesdropper acting on the quantum channel, derive its theoretical bounds, and show  that the achievable SKR closely approaches the upper limit, yielding a substantial improvement over RR based exclusively on hard information (RRH for short);
\item We derive a lower bound for the SKR against collective attacks when an eavesdropper is present on the quantum channel.
    Our SKR bound corresponds to a modified version of the Devetak--Winter bound \cite{devetak2005distillation},
    where the proposed reconciliation procedure solely improves the MI term between Alice and Bob, leaving the Holevo term between Bob and Eve unchanged;
\item We validate the proposed approach through a coded-level implementation using LDPC codes and belief-propagation decoding with 4-PAM and 8-PAM constellations, and assess its performance through numerical simulations, confirming its advantage over RRH;
\item We show the advantage of RRS over RRH in terms of SKR in the scenario of collective attacks on a linear quantum channel.
\end{itemize}
The proposed reconciliation scheme involves Bob disclosing a soft metric that enables Alice to perform soft decoding, without revealing any additional information to Eve beyond the syndrome normally required for error correction.

The paper is organized as follows. After providing a brief overview of the proposed reconciliation scheme in Section \ref{sec:overview}, we dive deep into the formal aspects of the problem in Section \ref{sec:theory},
presenting a solution in Section \ref{sec:transformation-functions}. 
In Section \ref{sec:simulations-results}, we investigate through numerical simulations the performance of the proposed solution. Finally, we draw some conclusions in Section \ref{sec:conclusions}.

\section{Reconciliation Scheme Overview}\label{sec:overview}

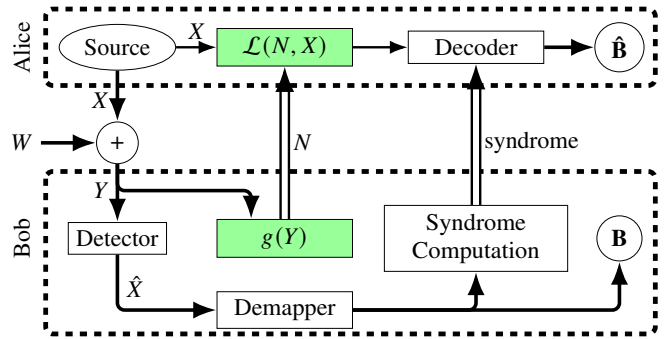
\begin{figure}[bt]
  \centering
  \resizebox{\columnwidth}{!}{%
  \begin{tikzpicture}[x=1em, y=1em, baseline=(current bounding box.north)]
\usetikzlibrary{arrows.meta}
\usetikzlibrary{shapes}
\newdimen\yoffset
\newdimen\ydemapper
\newdimen\xoffsetreconciliation
\newdimen\xoffsetsyndrome
\newdimen\xoffsetB
\setlength{\yoffset}{4em}
\setlength{\xoffsetreconciliation}{10em}
\setlength{\xoffsetsyndrome}{18em}
\setlength{\xoffsetB}{24em}

\draw[dashed, line width=2pt, rounded corners=3pt] (0, .6*\yoffset) rectangle (25.5em, 1.4*\yoffset);
\draw[dashed, line width=2pt, rounded corners=3pt] (0, -1.2em) rectangle (25.5em, -2*\yoffset);

% \draw[style=help lines] (0,-2*\yoffset) grid[step=(1em : 1em)] (50em,2*\yoffset);
% \draw (3em, \yoffset) ellipse (2.5em and 1em) node (source) {Source};
\node (source) [style={ellipse, draw}] at (3em, \yoffset) {Source};
\node[style={rectangle, draw}] (detector) at (3em, -\yoffset) {Detector} ;
%\node[anchor=center] at (3em, -\yoffset) (detector) {Detector};
\node [style={circle, draw}] (plus_circle) at (3em,0) {$+$};
\node (Z_text) at (-1em, 0em) {$W$};
\node [anchor=east] at (3em, 0.45*\yoffset) {$X$};
\node [anchor=east] at (3em, -0.5*\yoffset) {$Y$};

\path [draw, line width=1.5pt, -Latex] (source) to (plus_circle);
\path [draw, line width=1.5pt, -Latex] (Z_text) to (plus_circle);
\path [draw, line width=1.5pt, -Latex] (plus_circle) to (detector);

\node (noisemapper) [draw, style=rectangle, text width=5em, align=center, fill=green!40] at (\xoffsetreconciliation, -\yoffset) {$g(Y)$};
\node (noisedemapper) [draw, style=rectangle, text width=5em, align=center, fill=green!40] at (\xoffsetreconciliation, \yoffset) {$\mathcal{L}(N, X)$}; %{$\mathbf{g}^{-1}(\cdot)$};
\path [draw, line width=1pt, double distance=2.5pt, -Latex] (noisemapper.north) to (noisedemapper.south);
\node[anchor=west] at (\xoffsetreconciliation, 0) {$N$};
\path[draw, line width=1.5pt, rounded corners=3pt, -Latex] (plus_circle.south) |- (\xoffsetreconciliation/2, -\yoffset/2) -| (noisemapper.150);

\node (synenc) [draw, style=rectangle, text width=7em, align=center] at (\xoffsetsyndrome, -\yoffset) {Syndrome Computation};
\node (syndec) [draw, style=rectangle, text width=5em, align=center] at (\xoffsetsyndrome, \yoffset) {Decoder};
\path [draw, line width=1pt, double distance=2.5pt, -Latex] (synenc.north) to (syndec.south);
\node[anchor=west] at (\xoffsetsyndrome, 0) {syndrome};

% \node (lapprlabel) [anchor=south] at (\xoffsetreconciliation, \yoffset) {LAPPR$(X)$};
% \path [draw, -Latex, line width=1pt] (source.east) to (syndec.west);
% \node (xhatlabel) [anchor=south] at (\xoffsetreconciliation, -\yoffset) {$\hat{X}$};
% \path [draw, -Latex, line width=1pt] (detector.east) to (synenc.west);

\node [style={circle, draw}] (wordestimated) at (\xoffsetB,\yoffset) {$\mathbf{\hat{B}}$};
\node [style={circle, draw}] (word) at (\xoffsetB,-\yoffset) {$\mathbf{B}$};
% \path [draw, line width=1.5pt, -Latex] (word.north) to (wordestimated.south);

\path [draw, line width=1.5pt, -Latex] (syndec.east) to (wordestimated);
%\path [draw, line width=1.5pt, -Latex] (synenc.east) to (word.west);

% \node (demapper lappr) [style=rectangle, draw, text width=5em] at (\xoffsetreconciliation,\yoffset) {LAPPR($X$)};
\path [draw, line width=1pt, -Latex] (source.east) to (noisedemapper.west) ;% (demapper lappr.west) ;
\path [draw, line width=1pt, -Latex] (noisedemapper.east) % (demapper lappr.east) 
    to (syndec.west) ;
\node [anchor=south] at (6.3em, \yoffset) {$X$};

\node (demapper bob) [style=rectangle, draw, text width=5em, align=center] at (\xoffsetreconciliation, -\yoffset - 3em) {Demapper};
\path [draw, line width=1.5pt, -Latex, rounded corners=3pt] (detector.south) |- (demapper bob.west);
\path [draw, line width=1.5pt, -Latex, rounded corners=3pt] (demapper bob.east) -| (synenc.south);
\path [draw, line width=1.5pt, -Latex, rounded corners=3pt] (demapper bob.east) -| (word.south);
\node [anchor=west] at (3em, -\yoffset-2em) {$\hat{X}$};

\node [rotate=90] at (-1em, \yoffset) {Alice};
\node [rotate=90] at (-1em, -\yoffset) {Bob};
\end{tikzpicture}%
  }%
  \caption{System overview of the scheme. The highlighted blocks enable the reverse reconciliation with soft information. Bob generates a soft metric $N$ from the symbol $Y$ he received
  and Alice uses it to generate the log-\emph{a posteriori} probability ratios ($\mathcal{L}(N,X)$) of the corresponding bits for the subsequent syndrome-based error correction.}\label{fig:system-overview}
\end{figure}

\figref{fig:system-overview} illustrates the working principle of a system employing the proposed RRS procedure%
, and Table~\ref{tab:summary-steps} summarizes its steps. In the initial signaling phase, Alice transmits random symbols $X$ to Bob over an AWGN channel (step~1).
Upon receiving  $Y$ as the channel output (step~2), Bob makes decisions, denoted as $\Xhat$ (step 2~(a)).
 
Next, the core phase of RRS begins. Based on the channel output $Y$, Bob computes a side-information variable $N$ and transmits it to Alice over a noiseless public side channel (step 2~(b)).
This variable is designed to provide Alice, who knows the transmitted symbols $X$, with soft information about Bob's decision $\Xhat$.
At the same time, $N$ is carefully constructed to reveal no information about $\Xhat$ to any potential eavesdropper who does not know $X$, as discussed in Section~\ref{sec:information-leakage}.
The specific transformation $N=g(Y)$ used to compute  this side information is detailed in Section~\ref{sec:transformation-functions}.

Alice then uses the side channel information $N$, along with her knowledge of the transmitted symbol $X$, to compute the log-\emph{a posteriori} probability ratios (LAPPRs) %
for the bits corresponding to Bob's decisions (step~3), as detailed in Section~\ref{sec:llr-construction}.

In the subsequent error correction phase, Alice and Bob use a previously agreed-upon error correction code (for simplicity, we assume here a binary code). After collecting a sufficient number of decisions $\XXhat$, Bob computes the syndrome %
of the binary sequence $\mathbf{B}$ corresponding to $\XXhat$ and sends it to Alice over the side channel (step~4).
Alice feeds the decoder with the syndrome and the LAPPRs to produce an estimate $\mathbf{\BBhat}$ of the binary sequence $\mathbf{B}$ (step~5).

Finally, privacy amplification (not shown in the scheme) is performed to remove any information potentially leaked through the syndrome and produce the final shared key.

\begin{table}[!tb]
    \caption{Summary of our reverse reconciliation scheme\\with soft information (RRS).}
    \label{tab:summary-steps}
    \vspace{-.75em}
    \centering
    
\begin{tabular}{lp{.8\columnwidth}}
    \hline
    1     & Alice randomly draws a symbol $X$ out of a possibly discrete constellation and sends it over the quantum channel.\\
    2     & Bob receives a noisy copy $Y$ of the transmitted symbol, and processes it:\begin{itemize}
    \item[(a)] He discretizes $Y$ to $\Xhat$, and he maps $\Xhat$ to a bit vector according to a labeling rule.
    \item[(b)] He applies a transformation function $g(Y)$ and he discloses the result $N$ over the public channel.
    \end{itemize}\\
    3     & Alice computes the LAPPRs $\mathcal{L}(X,N)$ associated to this bit vector, taking $N$ into account.\\\hline
    1--3  & Steps 1--3 are repeated a sufficient number of times, so that Bob has collected a binary vector $\mathbf{B}$, and Alice has computed the associated LAPPRs.\\\hline
    4     & Bob computes and discloses the syndrome of $\mathbf{B}$.\\
    5     & Alice uses the syndrome and the computed LAPPRs for coset decoding, obtaining $\mathbf{\hat{B}}$.\\
    \hline
\end{tabular}

\end{table}

\section{Theory}\label{sec:theory}
In this section, we focus our analysis on a PAM signaling system. The analysis can be readily applied to QAM signaling by considering independently the in-phase and quadrature PAM signal components.

Alice, the sender, selects a random sample $X$ from a PAM alphabet $\Abet=\{a_1, \ldots, a_{M}\}$ according to the probability mass function (PMF)%
\footnote{Throughout this work, we will denote the probability density function of a generic random variable $A$ %
with $f_A$, and the corresponding cumulative distribution function (CDF) with $F_A$. For PMFs, we will use \mbox{$\PPP[A]{a}:=\PP{A=a}$}.}
$\PPP[X]{a_i}$ and sends it over an AWGN channel.
This PMF is arbitrary, therefore the following analysis holds when PAS-QAM is employed \cite{notarnicola2023probabilistic,roumestan2024shaped}, even though we will use uniform input probabilities in our simulations as a proof of concept.
Similarly, the channel noise need not be Gaussian, as long as its distribution is known both to Alice and Bob.
The noise sample added by the channel is $W \sim \mathcal{N}\matchparen{0, \frac{N_0}{2}}$. The noise variance is supposed to be known by Alice and Bob, for instance after channel estimation. The channel output is $Y = X + W$, with conditional distribution 
\begin{equation}\label{eq:pdf_y_x}
f_{Y|X}(y|a_i) = \frac{1}{\sqrt{\pi N_0}} \exp\left(-\frac{(y - a_i)^2}{N_0}\right)
\end{equation}
and marginal distribution
\begin{equation}\label{eq:pdf_y}
    f_Y(y) = \frac{1}{\sqrt{\pi N_0}}\sum_{a_j \in \Abet}\PPP[X]{a_j} \exp\left(-\frac{(y - a_j)^2}{N_0}\right).
\end{equation}
Based on $Y$ and a set of intervals $\{D_i \subset \mathbb{R}: \; \bigcup_{i=1}^M D_i = \mathbb{R}, \; D_k\bigcap D_j=\emptyset\; \forall k\neq j\}_{i=1}^M$ that partition the set of real numbers,
Bob makes a decision on $\Xhat$
\begin{equation} \label{eq:decisions}
    \xhat = \begin{cases}
    a_1, & y \in D_1 \\
    \vdots & \\
    a_M, & y \in D_M.
  \end{cases}
\end{equation}
The choice of the intervals $D_i$ (or equivalently, the decision thresholds, corresponding to their infimum and supremum points) is arbitrary. For example, they may depend on the noise variance $N_0$, or they could be fixed.
Furthermore, even though in this work we assume that the size of the \emph{output} alphabet matches the size of the \emph{input} alphabet, %
the analysis of this reconciliation scheme still applies when input and output alphabets have different sizes.
When both $X$ and $\Xhat$ belong to the same alphabet $\Abet$, the intervals $D_i$  may be chosen, for instance, to maximize the \emph{a posteriori} probabilities of the transmitted symbol $X$ (MAP criterion).

Once a sufficient number of decisions have been collected, Bob demaps the resulting sequence $\XXhat$ to a bitstring $\BB$, e.g., assuming a Gray mapping.
Then he evaluates its syndrome %
according to a code previously agreed upon, and supposed to be publicly known.
The syndrome is sent to Alice, who will use it to perform the error correction processing and get $\BBhat$, possibly identical to $\BB$.

To assist Alice in decoding $\BBhat$, Bob provides her with additional side information derived from $Y$. Since this information is shared over a public channel, its design must ensure that no information about Bob's decisions $\Xhat$ is leaked to a potential eavesdropper unaware of $X$.

To achieve this, Bob applies a deterministic transformation to $Y$, generating a random variable $N$ that is sent to Alice. 
Without loss of generality, we define the transformation between $Y$ and $N$ through the piecewise function
\begin{equation}\label{eq:functions}
      n = g(y) = \begin{cases}
        g_1(y), & y \in D_1\\
        \vdots & \\
        g_M(y), & y \in D_M.
    \end{cases}
\end{equation}

When the eavesdropper has no side information, the achievable SKR is $I(X;Y)$ \cite[Prop.~1]{ahlswede_common_1993}, \cite[Th.~2]{maurer_secret_1993}.
In the DR case, the raw key coincides with the input sequence $\XX$ transmitted by Alice, and Bob can exploit the continuous (soft) outputs $\YY$ to recover it. This reduces to the classical problem of reliable communication over a discrete-input AWGN channel, whose capacity can be closely approached with well-established techniques (e.g., Alice discloses the syndrome of $\XX$ based on an LDPC code, and Bob uses it with soft information from $\YY$ in a belief-propagation decoder).

In the RR case---the most relevant for QKD---Bob derives the raw key%
\footnote{%
In contrast to MDR, where the raw key is locally generated by Bob,
in RRS it is derived from a discretization of the channel output.
MDR employs the channel output to perform multidimensional rotations on a transformed version
of the raw key to produce a soft metric, whereas in RRS the soft metric is directly derived from the channel output,
as we will explain shortly.%
} %
$\XXhat$ from $\YY$, while Alice only has hard information (her sequence $\XX$) to recover it. Proceeding symmetrically to the DR case, i.e., with Bob sharing only the syndrome of $\XXhat$ and Alice using it with $\XX$ to recover $\XXhat$, the achievable rate is limited to $I(\Xhat;X)$, with significant loss compared to the ultimate rate $I(X;Y)$. 
If, however, Bob publicly discloses additional side information $N$, as in the RRS scheme of \figref{fig:system-overview}, the achievable SKR becomes $I(\Xhat;X|N)$ \cite[Th.~3]{ahlswede_common_1993}. This can be higher or lower than $I(\Xhat;X)$, depending on whether $N$ reveals more (or less) information to Alice than to Eve. We will refer to $I(\Xhat;X)$ and $I(\Xhat;X|N)$ as the achievable SKRs for the RRH and RRS schemes, respectively. 
In QKD, further side information may be gathered by Eve by tampering with the quantum channel, and must therefore be included in the computation of the SKR \cite{devetak2005distillation}.

\subsection{Constraint on Information Leakage}\label{sec:information-leakage}
A well-designed transformation $g$, according to the discussion above, should maximize $I(\Xhat;X|N)$.
First note that
\begin{equation}\label{eq:cond-mi}
    I(\Xhat;X|N) = I(\Xhat;X,N) - I(\Xhat;N).
\end{equation}
Whereas the optimal transformation $g$ may be hard to devise, it is reasonable to first minimize the second term $I(\Xhat;N)$, which represents the information about the raw key $\mathbf{B}$ leaked through the disclosed metric $N$. As the MI is non-negative,
we can reasonably impose the constraint
\begin{equation}\label{eq:mutual-information-null}
  I(\Xhat;N) = 0.
\end{equation}
The mutual information between two random variables is zero if and only if they are independent. Therefore, to satisfy (\ref{eq:mutual-information-null}) the function pieces $g_i$ in (\ref{eq:functions}) must be selected such that, given the decision $\Xhat=a_i$, the 
distribution of the resulting variable $N$ is independent of the decision $a_i$ itself, i.e.,
\begin{equation}\label{eq:conditional-distribution-all-equal}
    f_{N|\hat{X}}(n|a_i) = f_{N|\hat{X}}(n|a_j) = f_N(n) \quad \forall\, a_i,\,a_j \in \mathcal{A}.
\end{equation}

Imposing \eqref{eq:mutual-information-null} does not guarantee the optimality of the solution.
However, as we will see in the results, the remaining degrees of freedom allow us to closely approach the upper bound $I(X;Y)$ as discussed in the following section.

\subsection{Bounds on the SKR $I(\Xhat;X|N)$}\label{sec:bounds}
It is useful to establish upper and lower bounds for the SKR of RRS.
\begin{equation}
    I(\Xhat;X) \leq I(\Xhat;X|N) \leq I(X;Y).
\end{equation}

\begin{IEEEproof}
    For the lower bound we first use \eqref{eq:mutual-information-null} in \eqref{eq:cond-mi} to get
    \begin{equation}\label{eq:cond-mi-equal-joint-mi}
        I(\Xhat;X|N) = I(\Xhat;X,N).
    \end{equation}
    From the chain rule and the non-negativity of MI
    \begin{equation}\label{eq:mi-bounds}
        I(\Xhat;X,N) = I(\Xhat;X) + I(\Xhat;N|X) \geq I(\Xhat;X).
    \end{equation}
    For the upper bound we have
    \begin{align}
    I(X;Y) & =    I(\Xhat, N; X) \label{lmiub:eq1} \\
           & =    I(\Xhat; X | N) + I(N; X) \label{lmiub:eq2}\\
           & \geq I(\Xhat ; X | N), \label{lmiub:eq3}
    \end{align}
    where
    \begin{itemize}
        \item \eqref{lmiub:eq1} follows from the fact that $Y$ and the pair $(\Xhat, N)$ are related by a deterministic invertible function;
        \item \eqref{lmiub:eq2} is the chain rule for mutual information;
        \item the inequality in \eqref{lmiub:eq3} is based on the non-negativity of mutual information.
    \end{itemize}
\end{IEEEproof}

\begin{remark*}
    The proof of the lower bound uses constraint \eqref{eq:mutual-information-null}.
    On the other hand, the proof of the upper bound does not, therefore it holds in general.
\end{remark*}

As expected, sharing $N$ over a public channel cannot increase the SKR beyond the limit imposed by \cite[Prop.~1]{ahlswede_common_1993}. At the same time, if $N$ satisfies \eqref{eq:mutual-information-null}, the SKR achievable by RRS is not lower than the SKR achievable by RRH.

\subsection{SKR with a Quantum Eavesdropper}\label{sec:skr-quantum}
In this section we analyze how the presence of an attacker, namely Eve,
would impact the computation of the SKR under the framework of collective attacks
and infinite key length.
In this scenario, the maximum
achievable SKR $K$ can be lower bounded as
\begin{equation}\label{eq:skr-bound}
    K \geq I(\Xhat;X|N) - \chi\left(Y;E\right),
\end{equation}
where $\chi$ is the Holevo bound on the information
that Eve can extract \cite[p. 622]{cariolaro_quantum_2015}.
\begin{IEEEproof}
    The amount of secret information that can be extracted by Alice and Bob,
    provided that the disclosed metric $N$ is known both by Alice and by Eve, is
    \cite{leverrier_multidimensional_2008,devetak2005distillation,Pirandola:20}
    \begin{equation}\label{eq:skr-complete}
        K = I(\Xhat;X,N) - \vnentr{\Xhat;E,N},
    \end{equation}
    where $E$ is a random variable that summarizes the eavesdropper's
    collected information.
    Since $\Xhat$ and $N$ are independent classical random variables,
    it follows that \cite[Lemma 1]{leverrier_multidimensional_2008}
    \begin{equation}\label{eq:inequality-lemma-1-leverrier}
        \vnentr{\Xhat;E,N}\leq\vnentr{\Xhat,N;E}.
    \end{equation}
    Since $g$ is invertible in each region $D_i$, each
    pair $(\Xhat, N)$ is statistically equivalent to $Y$. Then,
    \begin{equation}\label{eq:vnentropy-equivalence}
        \vnentr{\Xhat,N;E} = \vnentr{Y;E}.
    \end{equation}
    Putting \eqref{eq:skr-complete}, \eqref{eq:inequality-lemma-1-leverrier},
    and \eqref{eq:vnentropy-equivalence} together, we can conclude that
    \begin{align}
        \label{eq:skr-inequality}
        K &\geq I(\Xhat;X,N) - \chi\left(Y;E\right)\\
        &=I(\Xhat;X|N) - \chi\left(Y;E\right),
    \end{align}
    where the last equality follows from \eqref{eq:cond-mi} and \eqref{eq:mutual-information-null}.
\end{IEEEproof}
This result implies that the soft metric we introduced impacts the MI between Alice and Bob, thus the reconciliation efficiency, and it does not affect the Holevo bound.

\subsection{Transformation Functions}\label{sec:transformation-functions}
To construct a solution to (\ref{eq:conditional-distribution-all-equal}), we begin by arbitrarily specifying the target output distribution $f_N(n)$ as the uniform distribution over the interval $[0,1]$. To preserve as much information as possible about the channel output $Y$ and ease the analytical calculation of the LAPPRs (Section \ref{sec:llr-construction}), we further require each $g_i$ to be continuously differentiable and invertible (therefore monotonic) in its domain $D_i$.

We are left with an additional degree of freedom for each function $g_i$, i.e., the direction of monotonicity---increasing or decreasing. We first consider the case in which all functions are monotonically increasing. A generalization to arbitrary monotonicity configurations is provided in
Section \ref{sec:monotonicity-configs}.

It is straightforward to verify 
that, conditioned on $Y\in D_i$, the transformation

\begin{equation}\label{eq:gi}
    g_i(y) = F_{Y|\Xhat}(y|a_i) =  \frac{F_Y(y) - F_Y(\inf D_i)}{\PPP[\Xhat]{a_i}},
\end{equation}
corresponding to the conditional CDF of $Y$ given $\Xhat=a_i$, produces a uniformly distributed output over the interval $[0,1]$.\footnote{A formal proof follows from the probability integral transform \cite[Th. 2.1.10]{casella_statistical_2002}.} Here, $F_Y(y)=\int_{-\infty}^y f_Y(z)dz$ is the CDF of the channel output $Y$, whose distribution $f_Y(y)$ is given in \eqref{eq:pdf_y}; {$\PPP[\Xhat]{a_i} = F_Y(\sup D_i) - F_Y(\inf D_i)$} is the probability that $Y\in D_i$, i.e., that Bob makes the decision $\Xhat=a_i$; and, for conciseness, we define 
\begin{equation}
    F_Y(\inf D_i) = \lim_{\tilde{y} \rightarrow \inf D_i}F_Y(\tilde{y}),
\end{equation}
and similarly for $F_Y(\sup D_i)$. %
Consequently, defining the global transformation $g(y)$ in (\ref{eq:functions}) by assigning each $g_i$ as in (\ref{eq:gi}) for $i=1,\ldots,M$, satisfies the constraint in~(\ref{eq:conditional-distribution-all-equal}).
In other words, the conditioned random variables \mbox{$N_i := N|\{\Xhat=a_i\}$} are all statistically equivalent, 
i. e.,
\mbox{$\forall i,j: \; N_i \cong N_j \cong N \sim \mathcal{U}\matchparen{\left[0,1\right]}$}.
Moreover, each function $g_i$ in (\ref{eq:gi}) %
is invertible and differentiable over its domain $D_i$, with
\begin{align}
  &g_i^{-1}(n) = F_Y^{-1}\left(n \cdot \PPP[\Xhat]{a_i} + F_Y(\inf D_i)\right)\label{eq:invgi},\\
  &g_i'(y) = \frac{f_Y(y)}{\PPP[\Xhat]{a_i}}.\label{eq:derivative}
\end{align}

\subsection{Secret Key Rate}\label{sec:mutual-information}

We can assess the performance of RRS by its SKR, or equivalently by the mutual information $I(\Xhat;X,N)$ (as proved earlier, see \eqref{eq:cond-mi-equal-joint-mi}):
\begin{equation}\label{eq:mi-expanded}
    I(\Xhat;X,N) = H(\Xhat) - H(\Xhat | X, N),
\end{equation}
where $H(\Xhat|X, N) = H(\Xhat, N|X) - H(N|X)$ can be written as
\begin{equation}\label{eq:cond-entropy}
    H(\Xhat|X, N) := \mathop{\mathbb{E}}_{\Xhat,X,N}\left\{\log_2\matchparen{\frac{
        \sum_{a_k\in\Abet} f_{N,\Xhat|X}(N, a_k | X)
    }{
        f_{N,\Xhat|X}(N, \Xhat | X)
    }}\right\},
\end{equation}
which is entirely expressed in terms of $f_{N,\Xhat|X}$.
Since the event \mbox{$\{N=n, \Xhat=a_i\}$} corresponds to the event \mbox{$\{Y=g_i^{-1}(n)\}$}, this distribution can be obtained through a change of variable as \cite[Th.~2.1.8]{casella_statistical_2002}
\begin{equation}\label{eq:n-xhat-cond-x}
    f_{N,\Xhat|X}(n, a_i|a_j) = \frac{
        f_{Y|X}(g_i^{-1}(n)|a_j)
    }{
        |g_i'(g_i^{-1}(n))|
    }.
\end{equation}

We remark that, so far, we have at no point assumed a particular distribution of the noise. We just assumed that it is known.
In the case of Gaussian noise distribution, the above distribution can be expressed as
\begin{equation}
    f_{N,\Xhat|X}(n, a_i|a_j) = \frac{
        \PPP[\Xhat]{a_i}
    }{
        \sum_{a_k\in\Abet}\PPP[X]{a_k}e^{-\frac{(2g_i^{-1}(n) - a_j - a_k)(a_j - a_k)}{N_0}}
    },
\end{equation}
where we replaced the expressions for $g_i'(y)$ from \eqref{eq:derivative}, for $f_Y(y)$ from \eqref{eq:pdf_y}, and for $f_{Y|X}$ from \eqref{eq:pdf_y_x}.
\subsection{A Posteriori Probability Ratios}\label{sec:llr-construction}
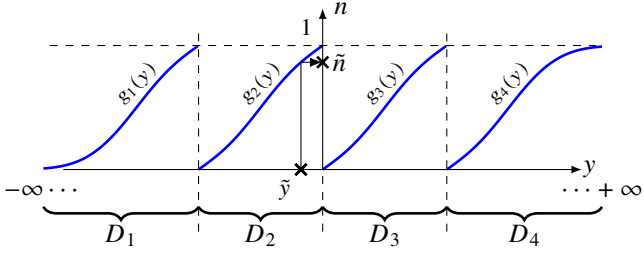
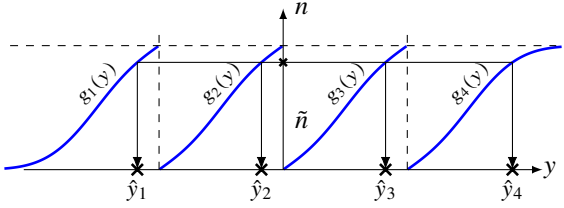
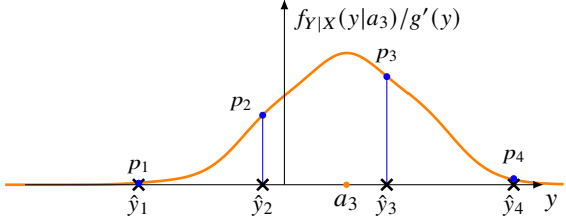
\begin{figure}[!tb]
    \centering
    \hfil\subfloat[Bob receives $y=\tilde{y}$, so he decides on $\xhat = a_2$, applies $g_2$ to compute $\tilde{n} = g_2(\tilde{y})$, and sends it to Alice.%
    ]{
        \hfil
        \begin{tikzpicture}
            \begin{axis}[
                    axis lines=none,
                    ymin=-1, ymax=1.4,
                    xmin=-4.5, xmax=4.5,
	            clip=false,
                    line width=1pt,
                    width=\columnwidth,
                    x={\xunitvect},
                    y=\yunitvect
                ]

                \addplot table [
                    mark=none,
                    col sep=comma,
                    unbounded coords=jump
                ] {./data_plot_function_8.000.csv};

                \draw[thin, -latex] (axis cs:-4.18, 0)  -> (axis cs:4.18, 0); %
                \draw[thin, -latex] (axis cs:0, 0)      -> (axis cs:0, 1.3); %

                \draw[thin, dashed] (axis cs:-2,-.5) -- (axis cs:-2,1.1);
                \draw[thin, dashed] (axis cs:2,-.5) -- (axis cs:2,1.1);
                \draw[thin, dashed] (axis cs:0,-.5) -- (axis cs:0,0);

                \draw[thin, dashed] (axis cs:-4.4,1) -- (axis cs:4.4,1);

                \node at (axis cs:4.3,0) {$y$};
                \node at (axis cs:0.3,1.3) {$n$};
                \node[anchor=south east] at (axis cs:0,1) {\small $1$};

                \node[rotate=45] at (axis cs: -3 , 0.7) {\scriptsize $g_{1}(y)$};
                \node[rotate=45] at (axis cs: -1 , 0.7) {\scriptsize $g_{2}(y)$};
                \node[rotate=45] at (axis cs:  1 , 0.7) {\scriptsize $g_{3}(y)$};
                \node[rotate=45] at (axis cs:  3 , 0.7) {\scriptsize $g_{4}(y)$};

                \draw[decorate, decoration={brace, amplitude=5pt}] (axis cs: 0, -.3) -- (axis cs: -2, -.3) node[midway, below, yshift=-3pt] {$D_2$};
                \draw[decorate, decoration={brace, amplitude=5pt}] (axis cs: 2, -.3) -- (axis cs: 0, -.3) node[midway, below, yshift=-3pt] {$D_3$};
                \draw[decorate, decoration={brace, amplitude=5pt, mirror}] (axis cs: -4.5, -.3) node[anchor=south] {$-\infty \cdots$} -- (axis cs: -2, -.3) node[midway, below, yshift=-3pt] {$D_1$};
                \draw[decorate, decoration={brace, amplitude=5pt}] (axis cs: 4.5, -.3) node[anchor=south] {$\cdots +\infty$} -- (axis cs: 2, -.3) node[midway, below, yshift=-3pt] {$D_4$};

                \node [anchor=north east] at (axis cs:\myytwo,  0) {\small $\tilde{y}$};
                \addplot [black, mark=x, mark size=3pt] coordinates {(\myytwo, 0    )};

                \node[anchor=west] at (axis cs:0, \myN) {$\tilde{n}$};
                \addplot [only marks, mark=x, mark color=black, mark size=3pt] coordinates {
                    (\myytwo, 0    )
                    (0   , \myN)
                };
                
                \draw[-Latex, thin] (axis cs:\myytwo, 0) |- (axis cs:0, \myN);
                
            \end{axis}
        \end{tikzpicture}%
        \hfil
        \label{fig:function-usage-bob}
    }\hfil\\\hfil
    \subfloat[%
            Alice (and Eve) receives $\tilde{n}$ from Bob and applies the inverse transformation function within each region to infer the possible channel outputs $\hat{y}_i=g_i^{-1}(\tilde{n})$ that Bob may have observed.%
    ]{
        \centering
        \begin{tikzpicture}
            \begin{axis}[
	            axis lines=none,
	            ymin=-.5, ymax=1.4,
                    xmin=-4.5, xmax=4.5,
                    line width=1pt,
                    width=\columnwidth,
                    clip=false,
                    x={\xunitvect},
                    y=\yunitvect
                ]
                
                \addplot [
                    blue
                ] table [
                    mark=none,
                    col sep=comma,
                    unbounded coords=jump
                ] {./data_plot_function_reduced.csv};

                \draw[thin, -latex] (axis cs:-4.18, 0)  -> (axis cs:4.18, 0); %
                \draw[thin, -latex] (axis cs:0, 0)      -> (axis cs:0, 1.3); %

                \draw[thin, dashed] (axis cs:-2,0) -- (axis cs:-2,1.1);
                \draw[thin, dashed] (axis cs:2,0) -- (axis cs:2,1.1);

                \draw[thin, dashed] (axis cs:-4.4,1) -- (axis cs:4.4,1);

                \node at (axis cs:4.3,0) {$y$};
                \node at (axis cs:0.3,1.3) {$n$};
                
                \node[rotate=45] at (axis cs: -3 , 0.7) {\scriptsize $g_{1}(y)$};
                \node[rotate=45] at (axis cs: -1 , 0.7) {\scriptsize $g_{2}(y)$};
                \node[rotate=45] at (axis cs:  1 , 0.7) {\scriptsize $g_{3}(y)$};
                \node[rotate=45] at (axis cs:  3 , 0.7) {\scriptsize $g_{4}(y)$};

                \node at (axis cs:0.3, 0.4) {$\tilde{n}$};
                \addplot [black, mark=x, mark size=2pt] coordinates { (0, \myN) };

                \addplot [only marks, mark=x, mark color=black, mark size=3pt] coordinates {
                    (\myyone, 0    )
                    (\myytwo, 0    )
                    ( \myythree, 0    )
                    ( \myyfour, 0    )
                };

                \draw[thin] (axis cs: \myyone, \myN) -- (axis cs: \myyfour, \myN);
                
                \draw[Latex-, thin] (axis cs:\myyone  , 0) -- (axis cs:\myyone  , \myN);
                \draw[Latex-, thin] (axis cs:\myytwo  , 0) -- (axis cs:\myytwo  , \myN);
                \draw[Latex-, thin] (axis cs: \myythree  , 0) -- (axis cs: \myythree  , \myN);
                \draw[Latex-, thin] (axis cs: \myyfour  , 0) -- (axis cs: \myyfour  , \myN);

                \node [anchor=north] at (axis cs:\myyone, 0) {\small$\hat{y}_1$};
                \node [anchor=north] at (axis cs:\myytwo, 0) {\small$\hat{y}_2$};
                \node [anchor=north] at (axis cs:\myythree, 0) {\small$\hat{y}_3$};
                \node [anchor=north] at (axis cs:\myyfour, 0) {\small$\hat{y}_4$};

            \end{axis}
        \end{tikzpicture}
        \label{fig:function-usage-alice}
    }\hfil\\\hfil
    \subfloat[Since Alice knows what she transmitted, e.g. $x=a_3$, she can compute the \emph{a posteriori} probabilities of $\Xhat$ given $X$, by the samples $ p_i \propto \text{P}_{\Xhat | X, N}(a_i|a_3, \tilde{n})$ %
        of \eqref{eq:n-xhat-cond-x} (plotted here as a function of $y$) at $\hat{y}_i$.%
    ]{
        \centering
        \begin{tikzpicture}
            \begin{axis}[
                    axis lines=none,
                    ymin=-0.5, ymax=1,
                    xmin=-4.5, xmax=4.5,
                    line width=1pt,
                    width=\columnwidth,
                    x=\xunitvect,
                    y={3*\xunitvect}
                ]

                \addplot [orange] table [
                    mark=none, col sep=comma,
                    x index=0, y index=4
                ] {./data_plot_function_8.000.csv};

                \draw[thin, -latex] (axis cs:-4.18, 0)  -> (axis cs:4.18, 0); %
                \draw[thin, -latex] (axis cs:0, 0)      -> (axis cs:0, 1) node [
                    anchor=north west
                ] {\small $f_{Y|X}(y|a_3)/g'(y)$}; %

                \node[anchor=north] at (axis cs:4.3,0) {$y$};

                \node[below]  at (axis cs:  1 , 0) {$a_{3}$};
                
                \addplot [only marks, mark=x, mark color=black, mark size=3pt] coordinates {
                    (\myyone, 0    )
                    (\myytwo, 0    )
                    ( \myythree, 0    )
                    ( \myyfour, 0    )
                };

                \filldraw[thin, orange] (axis cs:  1 , 0) circle (1pt);
                
                \addplot [blue, thin, ycomb, mark=*, mark size=1.1pt] coordinates {
                    (\myyone,  \myfone)
                    (\myytwo, \myftwo+.03)
                    (\myythree, \myfthree-.025)
                    (\myyfour, \myffour-.03)
                };

                \node [anchor=north] at (axis cs:\myyone, 0) {\small$\hat{y}_1$};
                \node [anchor=north] at (axis cs:\myytwo, 0) {\small$\hat{y}_2$};
                \node [anchor=north] at (axis cs:\myythree, 0) {\small$\hat{y}_3$};
                \node [anchor=north] at (axis cs:\myyfour, 0) {\small$\hat{y}_4$};

                \node [anchor=south] at (axis cs:\myyone, \myfone) {\footnotesize $p_1$};
                \node [anchor=south east] at (axis cs:\myytwo, \myftwo) {\footnotesize $p_2$};
                \node [anchor=south] at (axis cs:\myythree, \myfthree) {\footnotesize $p_3$};
                \node [anchor=south] at (axis cs:\myyfour, \myffour) {\footnotesize$p_4$};
            \end{axis} ;
        \end{tikzpicture}
        \label{fig:function-usage-distribution}
    }\hfil
    \caption{Usage of the transformation function.\label{fig:function-usage}}
\end{figure}
To apply the RRS scheme described above, we finally detail the computation of the LAPPRs that Alice can feed to a syndrome-aware decoder to recover the raw key $\mathbf{\hat{B}}$.

For the sake of clarity, we will follow a concrete example, sketched in \figref{fig:function-usage}.
Suppose that Bob receives the channel output $\tilde{y} \in D_2$, as shown in \figref{fig:function-usage-bob}.
Therefore, he assigns $\xhat=a_2$ according to \eqref{eq:decisions}, and computes $\tilde{n}=g(\tilde{y})=g_2(\tilde{y})$ based on \eqref{eq:functions}. Finally, he publicly discloses $\tilde{n}$.

\figref{fig:function-usage-alice} shows that both Alice and Eve can now apply the set of locally inverse functions to $\tilde{n}$, and formulate $M$ (in our example, $M=4$) different hypotheses $\hat{y}_1,\ldots,\hat{y}_M$ for the channel output that Bob may have received, where $\hat{y}_i=g_i^{-1}(\tilde{n})$.
Note that the true channel output $\tilde{y}$ is included among these hypothetical values.

From Eve's perspective, $\tilde{n}$ alone provides no information about Bob's decision $\xhat$.
In fact, she could use a soft decoder with a soft input computed from $\PPP[Y|N]{\hat{y}_i|\tilde{n}} = \PPP[\Xhat|N]{a_i|\tilde{n}}$. However, such input reduces to the \emph{a priori} probability $\PPP[\Xhat]{a_i}$ due to \eqref{eq:conditional-distribution-all-equal} and the uniform distribution of $N$ over $[0,1]$.

On the other hand, Alice knows which symbol $x$ she transmitted.
Therefore, unlike Eve, she can combine this knowledge with $\tilde{n}$ to estimate and compare the \emph{a posteriori} probabilities of the $M$ possible decisions made by Bob.
In the example of \figref{fig:function-usage-distribution}, $x=a_3$, meaning that, due to channel noise, Alice's transmitted symbol differs from Bob's decision $\xhat=a_2$. However, by substituting $x=a_3$ into \eqref{eq:n-xhat-cond-x}, Alice obtains $f_{Y|X}(y|a_3)/g'(y)$ (the orange curve), and can evaluate it at $y=\hat{y}_i$ to derive the (unnormalized) \emph{a posteriori} probabilities $p_i\propto \PPP[\Xhat|X,N]{a_i|a_3, \tilde{n}}$ for $i=1,\ldots,4$. 
These probabilities indicate that, although her transmitted symbol $a_3$ remains the most likely, the decision $a_2$---Bob's actual choice---has a comparable probability, which may help Alice's decoder correct the discrepancy.

In general, the (normalized) \emph{a posteriori} probability required by Alice's decoder can be expressed as

\begin{equation}\label{eq:app-interval}
    \PPP[\Xhat|X,N]{a_i \; | \; a_j, \; n} =\\%
    \frac{%
        f_{N,\Xhat|X}(n, a_i| a_j)
    }{%
        f_{N|X}(n| a_j)
    }.
\end{equation}
For bitwise decoding, the LAPPR associated with the $l$-th bit $B_l$ of the label assigned to $\Xhat$ is
\begin{equation}\label{eq:llr}
  \begin{split}
    \mathcal{L}_l(n, a_j) := 
    &\log \matchparen {%
            \frac{
                \PPP[B_l|X,N]{0|a_j, n} %
            }{%
                \PPP[B_l|X,N]{1|a_j, n} %
            }%
    }\\
    =&\log \matchparen {%
            \frac{
                \sum_{a_i\in\Abet_0(l)} %
                    \PPP[\Xhat|X,N]{a_i \; | \; a_j, \; n} %
            }{%
                \sum_{a_i\in\Abet_1(l)} %
                    \PPP[\Xhat|X,N]{a_i \; | \; a_j, \; n} %
            }%
    }\\
    = &\log \matchparen {%
            \frac{
                \sum_{a_i\in\Abet_0(l)} %
                    f_{N,\Xhat|X}(n, a_i| a_j)
            }{%
                \sum_{a_i\in\Abet_1(l)} %
                    f_{N,\Xhat|X}(n, a_i| a_j)
            }%
    },
    \end{split}
\end{equation}
where $\Abet_0(l),\Abet_1(l)\subset\Abet$ partition $\Abet$ according to the value of the $l$-th bit (0 or 1, respectively),
based on a specific symbol-to-bit labeling rule---Gray labeling, for instance.

The denominator $f_{N|X}(n| a_j)$ in \eqref{eq:app-interval} ensures proper normalization of the probabilities. However, this term cancels out in the computation of the LAPPRs, as shown in the last equality of \eqref{eq:llr}. Consequently, Alice can directly use unnormalized probabilities; in the example of \figref{fig:function-usage-distribution}, she computes $\mathcal{L}_l(\tilde{n}, a_3)$ by using the samples $p_i$ shown in the figure.

\begin{remark*}
    Compared to RRH, RRS requires the additional computation of the soft metric by Bob
    and its inverse for the computation of the LAPPRs by Alice.
    In practice they involve the computation of functions of a single real variable that look like
    the plot of \figref{fig:function-usage}, which can be implemented with low complexity through lookup tables.
\end{remark*}
\setcounter{remark}{0}

\subsection{Direction of the Monotonicity}\label{sec:monotonicity-configs}
If $N_i$, as defined in Section \ref{sec:transformation-functions}, is uniformly distributed in $[0, 1]$, so is the random variable $\bar{N}_i = 1-N_i$. The latter can be obtained through the modified transformation
\begin{equation}
  \bar{g}_i(y) = \frac{F_Y(\sup D_i) - F_Y(y)}{\PPP[\Xhat]{a_i}} = 1 - g_i(y)\\
\end{equation}
corresponding to the complementary conditional CDF of $Y$ given $\Xhat=a_i$. By construction, $\bar{g}_i$ is also invertible and differentiable in its domain $D_i$, with

\begin{align}
  &\bar{g}_i^{-1}(n) = F_Y^{-1}\left(F_Y(\sup D_i) - n \cdot \PPP[\Xhat]{a_i}\right),\\
  &|\bar{g}_i'(y)| = -\bar{g}_i'(y) = \frac{f_Y(y)}{\PPP[\Xhat]{a_i}}.\label{eq:abs-derivative}
\end{align}

Since $\bar{N}_i$ has the same uniform distribution as $N_i$, Alice and Bob can agree on using either of them independently on each interval, 
i.e., if they agree to use $N_i$ for channel outputs that lie in $D_i$, they can arbitrarily agree to use either 
$N_j$ or $\bar{N}_j$ when $Y$ lies is another interval $D_j$. Either choice complies with constraint (\ref{eq:conditional-distribution-all-equal}) and is therefore still a valid solution to our problem, i.e., 
 no information about Bob's decision is obtained from observing $N$ alone.

On the other hand, the set of monotonicity directions chosen in the various intervals does impact the distinguishability of the hypothetical channel outputs from Alice's point of view. %

We define a \emph{configuration}  $C^{[b]}$ as an $M$-tuple of monotonicity directions, one for each decision interval. Each of the $2^M$ possible configurations is identified by a superscript $[b]$, where $b\in{0,1,\ldots,2^M-1}$. 
The monotonicity direction in each decision interval $D_i$ is characterized by the binary representation of $b$, i.e.,  $C^{[b]}=(b_M,\ldots,b_1)$: 
$b_i=0$ indicates that the transformation is monotonically increasing ($\shortarrow{1}$) in $D_i$; $b_i=1$ indicates that it is monotonically decreasing ($\shortarrow{-1}$).
For instance, for $M=4$ (4-PAM), we have $2^4=16$ different configurations, e.g., the "base" configuration%
, whose index has binary representation $C^{[0]}=(0,0,0,0)$, and the "alternating" configuration 
, whose index has binary representation $C^{[5]}=(0,1,0,1)$. Each configuration $C^{[b]}$ defines a different transformation $G^{[b]}$, which is a piecewise function over the decision intervals (analogously to \eqref{eq:functions})
\begin{equation}\label{eq:functions_G}
      n = G^{[b]}(y) = \begin{cases}
        G^{[b]}_1(y), & y \in D_1\\
        \vdots & \\
        G^{[b]}_M(y), & y \in D_M.
    \end{cases}
\end{equation}
where
\begin{equation}
      G^{[b]}_i(y) = \begin{cases}
        g_i(y) &   \text{if } b_i=0\\        
        1-g_i(y) & \text{if } b_i=1.
    \end{cases}
\end{equation}
For instance, the aforementioned "base" and "alternating" configurations, would visually look like
\begin{align}
    G^{[0]}&\sim[\shortarrow{1},\shortarrow{1},\shortarrow{1},\shortarrow{1}]\\
    G^{[5]}&\sim[\shortarrow{-1},\shortarrow{1},\shortarrow{-1},\shortarrow{1}].
\end{align}
The analysis of the SKR in Section \ref{sec:mutual-information} still applies, where $g_i$ is simply replaced by $G_i^{[b]}$ in \eqref{eq:n-xhat-cond-x}.

Some monotonicity configurations are \emph{equivalent}, in the sense that they yield identical SKRs when used in the RRS scheme.
As shown in appendix \ref{sec:appendix-configurations}, equivalent configurations also produce the same conditional distribution \eqref{eq:n-xhat-cond-x}. Therefore the LAPPR distribution remains unchanged, and the BER performance is identical across all equivalent configurations.

Recognizing equivalent configurations is particularly useful when assessing the performance of RRS, as it allows to restrict numerical computations and simulations to a single representative from each equivalence class.
For an exhaustive study of equivalence classes, refer to the Appendix.
\section{Simulations and Results}\label{sec:simulations-results}
\subsection{Secret Key Rate}\label{sec:simulations-results-mi}
To assess the performance of the proposed RRS scheme, we compare its achievable SKR $I(\Xhat;X|N)$ against the achievable SKR of the RRH scheme $I(\Xhat;X)$ and the SKR upper bound $I(X;Y)$.
Results are obtained through a numerical library we developed for this purpose \cite{origlia_2025_15222665}.

The $M$ decision intervals in \eqref{eq:functions_G} are defined by means of $M-1$ thresholds (the first and last intervals being lower and upper bounded by $-\infty$ and $+\infty$, respectively). In this work, we consider two possible threshold selection strategies:
\begin{itemize}
    \item[(F)] where thresholds are \emph{fixed} across different SNR values and placed midway between adjacent constellation points. For 4-PAM, with constellation points $\{-3,-1,1,3\}$, the thresholds are at $-2, 0, +2$;
    \item[(A)] where thresholds are \emph{adaptively} selected at each SNR value to produce uniform symbol probabilities after hard decision, thus maximizing the entropy $H(\Xhat)$.\footnote{Maximizing $H(\Xhat)$ does not guarantee the maximization of the SKR in \eqref{eq:mi-expanded}, which depends also on $H(\Xhat|X,N)$. In principle, a joint optimization of \eqref{eq:mi-expanded} over both the input probabilities $\text{P}_X(a_j)$ and the decision thresholds would be required to maximize the SKR, but this is beyond the scope of this work.}
\end{itemize}

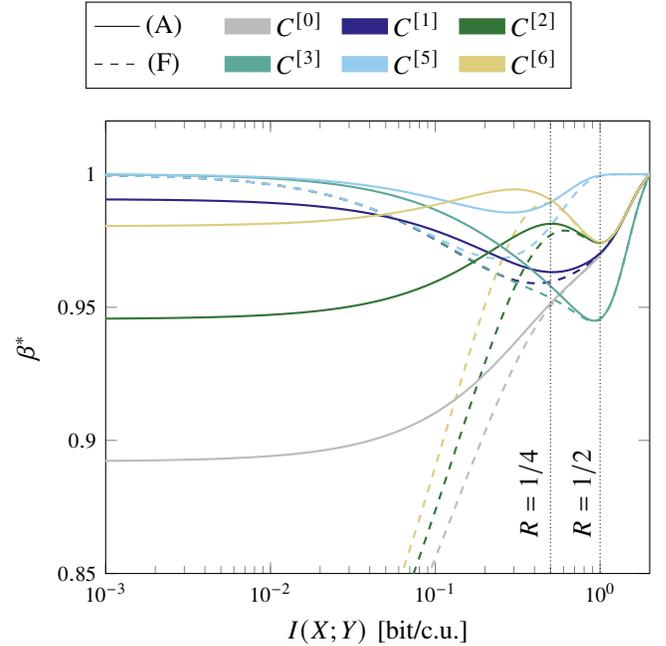
\begin{figure}[bt]
    \centering
    \ref{legend:MI-PAM4-config}\\
    \vspace{.2em}
    \begin{tikzpicture}
        \begin{semilogxaxis}[
            xmin=1e-3,
            ymin=.85,
            xmax=2,
            ymax=1.02,
            width=\columnwidth,
            legend to name={legend:MI-PAM4-config},
            legend columns=5,
            every axis legend/.append style={
                /tikz/every even column/.append style={
                    column sep=.5em
                }
            },
            xlabel={$I(X;Y)$ [bit/c.u.]},
            ylabel={$\beta^*$},
            every tick label/.append style={
                font=\ticklabelsize
            }%
        ]
            \addlegendimage{solid} \addlegendentry{(A)}
            \addlegendimage{empty legend}\addlegendentry{\hspace{3em}}
            \addlegendimage{area legend, fill, color=ptb1} \addlegendentry{$C^{[0]}$}
            \addlegendimage{area legend, fill, color=ptb2} \addlegendentry{$C^{[1]}$}
            \addlegendimage{area legend, fill, color=ptb3} \addlegendentry{$C^{[2]}$}
            \addlegendimage{dashed} \addlegendentry{(F)}
            \addlegendimage{empty legend}\addlegendentry{\hspace{3em}}
            \addlegendimage{area legend, fill, color=ptb4} \addlegendentry{$C^{[3]}$}
            \addlegendimage{area legend, fill, color=ptb5} \addlegendentry{$C^{[5]}$}
            \addlegendimage{area legend, fill, color=ptb6} \addlegendentry{$C^{[6]}$}
            
            \addplot [line width=0.8pt, dashed, forget plot, color=ptb1] table [
                col sep=comma,
                y expr=\thisrow{F0}/\thisrow{DIR},
                x=DIR
            ] {./data_pam4_mi_configurations.csv};

            \addplot [line width=0.8pt, dashed, forget plot, color=ptb2] table [
                col sep=comma,
                y expr=\thisrow{F1}/\thisrow{DIR},
                x=DIR
            ] {./data_pam4_mi_configurations.csv};

            \addplot [line width=0.8pt, dashed, forget plot, color=ptb3] table [
                col sep=comma,
                y expr=\thisrow{F2}/\thisrow{DIR},
                x=DIR
            ] {./data_pam4_mi_configurations.csv};

            \addplot [line width=0.8pt, dashed, forget plot, color=ptb4] table [
                col sep=comma,
                y expr=\thisrow{F3}/\thisrow{DIR},
                x=DIR
            ] {./data_pam4_mi_configurations.csv};

            \addplot [line width=0.8pt, dashed, forget plot, color=ptb5] table [
                col sep=comma,
                y expr=\thisrow{F5}/\thisrow{DIR},
                x=DIR
            ] {./data_pam4_mi_configurations.csv};

            \addplot [line width=0.8pt, dashed, forget plot, color=ptb6] table [
                col sep=comma,
                y expr=\thisrow{F6}/\thisrow{DIR}, 
                x=DIR
            ] {./data_pam4_mi_configurations.csv};
            
            \addplot [line width=0.8pt, solid, forget plot, color=ptb1] table [
                col sep=comma,
                y expr=\thisrow{A0}/\thisrow{DIR},
                x=DIR
            ] {./data_pam4_mi_configurations.csv};

            \addplot [line width=0.8pt, solid, forget plot, color=ptb2] table [
                col sep=comma,
                y expr=\thisrow{A1}/\thisrow{DIR},
                x=DIR
            ] {./data_pam4_mi_configurations.csv};

            \addplot [line width=0.8pt, solid, forget plot, color=ptb3] table [
                col sep=comma,
                y expr=\thisrow{A2}/\thisrow{DIR},
                x=DIR
            ] {./data_pam4_mi_configurations.csv};

            \addplot [line width=0.8pt, solid, forget plot, color=ptb4] table [
                col sep=comma,
                y expr=\thisrow{A3}/\thisrow{DIR},
                x=DIR
            ] {./data_pam4_mi_configurations.csv};

            \addplot [line width=0.8pt, solid, forget plot, color=ptb5] table [
                col sep=comma,
                y expr=\thisrow{A5}/\thisrow{DIR},
                x=DIR
            ] {./data_pam4_mi_configurations.csv};

            \addplot [line width=0.8pt, solid, forget plot, color=ptb6] table [
                col sep=comma,
                y expr=\thisrow{A6}/\thisrow{DIR}, 
                x=DIR
            ] {./data_pam4_mi_configurations.csv};

            \addplot [black, densely dotted, forget plot] coordinates {(.5, 0) (.5, 1.02)};
            \addplot [black, densely dotted, forget plot] coordinates {(1, 0) (1, 1.02)};

            \node [rotate=90, anchor=east] at (axis cs:.38, .9) {$R=1/4$};
            \node [rotate=90, anchor=east] at (axis cs:.8, .9) {$R=1/2$};
        \end{semilogxaxis}
    \end{tikzpicture}
    \vspace{-1em}
    \caption{Achievable reconciliation efficiency $\beta^*$  of the RRS scheme with 4-PAM, for different configurations $C^{[b]}$ and threshold selection strategies (A/F).}
    \label{fig:mi-configurations}
\end{figure}

We first consider the case of BPSK modulation ($M=2$), which admits only two non-equivalent configurations, $C^{[0]}$ and $C^{[1]}$ (see Table~\ref{tab:equivalent-configurations-bpsk} in Appendix \ref{appendix:table-equivalent-configs}), with the optimal threshold fixed at 0 by symmetry. In this setting, configuration $C^{[1]}$ corresponds to Leverrier's scheme \cite{leverrier2009theoretical}, where $|y|$ is remapped from $\mathbb{R}^+$ to $[0,1]$ to serve as $n$. This scheme was shown to be equivalent to the BI-AWGN channel, and therefore its achievable SKR attains the upper bound $I(X;Y)$. For this reason, the BPSK results are not explicitly plotted.

We then turn to the 4-PAM case. To better illustrate the performance of the proposed scheme, we introduce the \emph{achievable reconciliation efficiency} as the ratio between the achievable SKR and its upper bound\footnote{The reconciliation efficiency $\beta$ is commonly defined in the literature as the ratio between the actual rate achieved with a practical code and its upper bound. Here, instead, we add the superscript "*" and refer to $\beta^*$ as \emph{achievable} to emphasize that it is an information-theoretic metric, which does not account for the performance of a specific error-correction code.}
\begin{equation}
    \beta^* = \frac{I(\Xhat;X|N)}{I(X;Y)}.
\end{equation}

The selection of the configuration can be carried out by precomputing the MI 
for all configuration classes in a broad SNR range, and picking the optimal one for the SNR value of interest.
The number of distinct configuration classes is reported in Table~\ref{tab:order-classes}
of Appendix~\ref{appendix:table-equivalent-configs}. Even though it grows rapidly with the modulation order,
practical QAM-based CV-QKD systems typically employ up to a 64-QAM format (equivalent to a 8-PAM modulation on each quadrature),
and rarely exceed the 256-QAM modulation format, with 16 constellation points per real dimension.
For such modulation orders, the aforementioned exhaustive approach is still feasible.
Alternatively, one can always select the configuration with alternating monotonicities
in adjacent decision intervals ($C^{[5]}$ for 4-PAM, and $C^{[85]}$ for 8-PAM). Even though it is not always optimal,
experimental results indicate that it exhibits good achievable reconciliation efficiency also for higher modulation orders.
Further investigation is required.

\figref{fig:mi-configurations} shows $\beta^*$ for a representative from each configuration equivalence class of 4-PAM, listed in Table~\ref{tab:equivalent-configurations}, and for both fixed and adaptive thresholds.
At rate $R=1/2$ (and higher), the difference between the two threshold selection strategies is negligible (for high SNR, the fixed thresholds already yield almost uniform symbol probabilities), and the alternating configuration $C^{[5]}$ attains the highest efficiency ($\approx 1$). For lower rates, the choice of the threshold selection strategy becomes more important, and the adaptive strategy is always more efficient. At rate $R=1/4$, $C^{[5]}$ with adaptive thresholds still achieves the highest efficiency, on par with $C^{[6]}$ and only slightly below 1. Indeed, $C^{[5]}$ is the best configuration for almost any rate (except for the range 0.1--0.25, where $C^{[6]}$ has a slightly higher efficiency) and is asymptotically optimal for both low and high rates. For this reason, from now on we will only show results for configuration $C^{[5]}$ with adaptive thresholds.

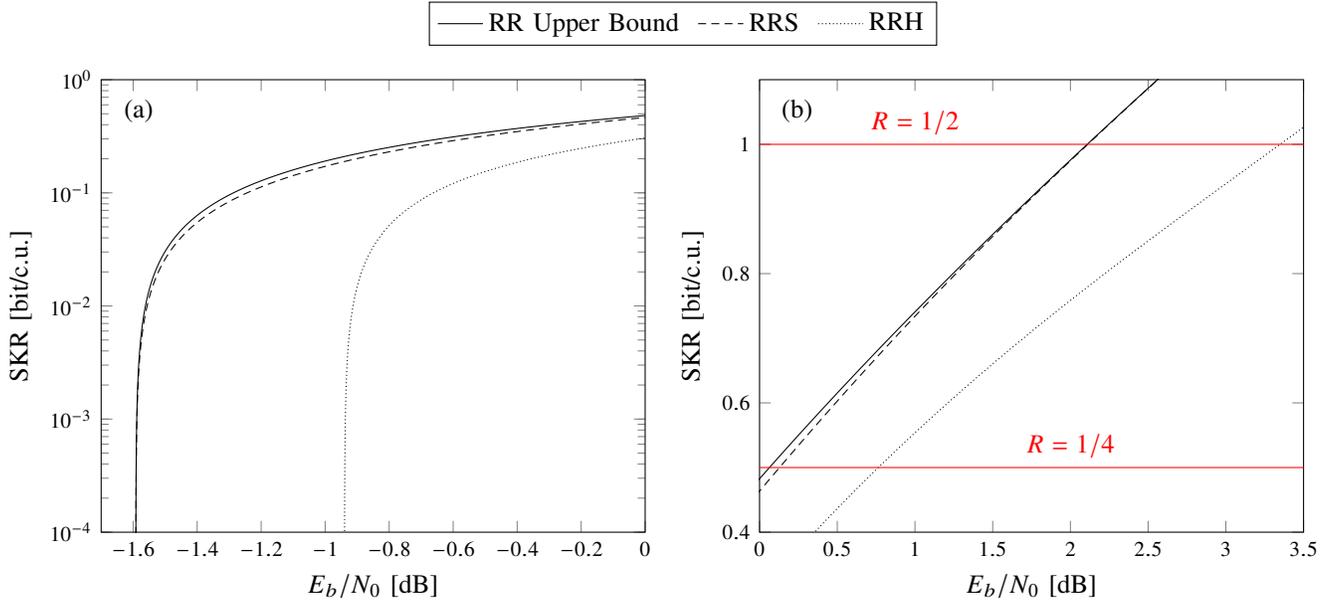
\begin{figure*}%
    \begin{center}
        \ref{legend:MI-PAM4}
    \end{center}
    
    \subfloat{
        \label{fig:mutual-information-pam4-all}
        \begin{tikzpicture}%
            \begin{semilogyaxis}[
                    xmin=-1.7,
                    ymin=1e-4,
                    xmax=0,
                    ymax=1,
                    width=\columnwidth,
                    legend to name={legend:MI-PAM4},
                    legend columns=5,
                    every axis legend/.append style={
                        /tikz/every even column/.append style={
                            column sep=.5em
                        }
                    },
                    xlabel={$E_b/N_0$ [dB]},
                    ylabel={SKR [bit/c.u.]},
                    every tick label/.append style={
                        font=\ticklabelsize
                    }%
                ]
                \addplot [black, no marks, dirls, dir colour] table [
                    col sep=comma,
                    x=scSNR, 
                    y=DIR
                ] {./data_mi_pam4_tcom.csv}; \addlegendentry{RR Upper Bound}\label{plot:DR}
                
                \addplot [black, no marks, rrsls, rrs colour] table [
                    col sep=comma,
                    x=scSNR_C5_U, 
                    y=C5_U
                ] {./data_mi_pam4_tcom.csv}; \addlegendentry{RRS}\label{plot:RRS}

                \addplot [black, rrhls, rrh colour] table [
                    col sep=comma,
                    x=scSNR_H-U, 
                    y=H-U
                ] {./data_mi_pam4_tcom.csv}; \addlegendentry{RRH}\label{plot:RRH}
            \end{semilogyaxis}
            \node at (.5,5.6) {(\subref*{fig:mutual-information-pam4-all})};
        \end{tikzpicture}
    }~%
    \subfloat{
        \label{fig:mutual-information-pam4-zoom}
        \begin{tikzpicture}%
            \begin{axis}[
                    xmin=0,
                    xmax=3.5,
                    ymin=.4,
                    ymax=1.1,
                    width=\columnwidth,
                    xlabel={$E_b/N_0$ [dB]},
                    ylabel={SKR [bit/c.u.]},
                    every tick label/.append style={
                        font=\ticklabelsize
                    }%
                ]
                \addplot [black, no marks, dirls, dir colour] table [
                    col sep=comma,
                    x=scSNR, 
                    y=DIR
                ] {./data_mi_pam4_tcom.csv};

                \addplot [black, no marks, rrsls, rrs colour, forget plot] table [
                    col sep=comma,
                    x=scSNR_C5_U, 
                    y=C5_U
                ] {./data_mi_pam4_tcom.csv};

                \addplot [black, rrhls, rrh colour] table [
                    col sep=comma,
                    x=scSNR_H-U, 
                    y=H-U
                ] {./data_mi_pam4_tcom.csv};

                \draw (axis cs: 0,1) --(axis cs: 1,1) node[above]{$R=1/2$}--(axis cs: 3.5,1);
                \draw (axis cs: 0,.5)--(axis cs: 2,.5)node[above]{$R=1/4$}--(axis cs: 3.5,.5);
            \end{axis}
            
            \node at (.5,5.6) {(\subref*{fig:mutual-information-pam4-zoom})};
        \end{tikzpicture}
    }
    \caption{Achievable SKR for 4-PAM with different reconciliation schemes and adaptive thresholds: %
        (\protect\subref*{fig:mutual-information-pam4-all}) %
        low-rate region in log scale; %
        (\protect\subref*{fig:mutual-information-pam4-zoom}) %
        region of interest for our BER simulations employing codes with rates \mbox{1/2} and \mbox{1/4}.%
    }\label{fig:mutual-information-pam4}
\end{figure*}

\figref{fig:mutual-information-pam4} shows the achievable SKR of the RRS and RRH schemes (both with same adaptive thresholds) compared with the upper bound $I(X;Y)$, as a function of the SNR $E_b/N_0$. 
For RRH, the minimum $E_b/N_0$ is about %
\num{-0.95}\unit{\dB} (\ref{plot:RRH}), with a gap of about \num{0.64}\unit{\dB} towards the upper bound (\ref{plot:DR}). At higher rates (shown in \figref{fig:mutual-information-pam4-zoom}), the gap increases to more than \num{1}\unit{\dB}. 
Consistently with the efficiency curve shown in \figref{fig:mi-configurations}, the RRS scheme (\ref{plot:RRS}) nearly closes this gap, tightly approaching the upper bound at both low and high rates, and showing just a small gap at intermediate rates. %

    To illustrate the advantage of RRS over RRH also in the presence of an eavesdropper performing a collective attack,
    we computed the corresponding secret key rate as a function of the distance, following the approach described in \cite{parente_discrete-modulation_2026}.
    We assumed a linear quantum channel model \cite{zhang_continuous-variable_2024,notarnicola_quantum_2025,parente_probabilistic_2025}, and a standard optical fiber with 0.2dB/km loss.
    The results are reported in \figref{fig:skr-collective} for excess noise values of $\xi=0.01$ and $\xi=0.03$ shot noise units.
    In both cases, RRS (\ref{plot:skr-rrs1}/\ref{plot:skr-rrs3}) yields a higher SKR and reaches
    a longer range then RRH (\ref{plot:skr-rrh1}/\ref{plot:skr-rrh3}), tightly approaching the upper bound when employing
    4-PAM transmissions with uniform input probabilities (\ref{plot:skr-dir1}/\ref{plot:skr-dir3}).
    Further improvements towards the GG02 bound are possible
    by applying probabilistic shaping of the constellation.

\begin{figure}[!tb]
    \centering
    \begin{tikzpicture}
        \begin{semilogyaxis}[
                ymin=1e-7, ymax=2,
                xmin=0, xmax=300,
	        clip=true,
                legend columns=2,
                legend style={
                    cells={anchor=west}
                },
                every axis legend/.append style={
                    /tikz/every even column/.append style={
                        column sep=.5em
                    }
                },
                filtered plot/.style={
                    x filter/.append expression={(x>180 && x<190) ? nan : \pgfmathresult},
                    x filter/.append expression={(x>190 && x<200) ? nan : \pgfmathresult},
                    x filter/.append expression={(x>240 && x<250) ? nan : \pgfmathresult},
                    x filter/.append expression={(x<10)           ? nan : \pgfmathresult},
                    x filter/.append expression={(x>40  && x<50 ) ? nan : \pgfmathresult},
                    x filter/.append expression={(x>50  && x<60 ) ? nan : \pgfmathresult},
                    x filter/.append expression={(x>90  && x<100) ? nan : \pgfmathresult},
                    x filter/.append expression={(x>100 && x<110) ? nan : \pgfmathresult}
                },
                width=\columnwidth,
                xlabel={Distance [km]},
                ylabel={SKR [bit/c.u.]}
            ]

            \addlegendimage{empty legend};\addlegendentry{\hspace{-2em}\color{T-Q-V2}$\xi=0.01$:};
            \addlegendimage{empty legend};\addlegendentry{\hspace{-2em}\color{T-Q-V4}$\xi=0.03$:};
            \addlegendimage{rrsls, rrs mark , xi 0.01};\addlegendentry{RRS};\label{plot:skr-rrs1}
            \addlegendimage{rrsls, rrs mark*, xi 0.03};\addlegendentry{RRS};\label{plot:skr-rrs3}
            \addlegendimage{rrhls, rrh mark , xi 0.01};\addlegendentry{RRH};\label{plot:skr-rrh1}
            \addlegendimage{rrhls, rrh mark*, xi 0.03};\addlegendentry{RRH};\label{plot:skr-rrh3}
            \addlegendimage{dirls, dir mark , xi 0.01};\addlegendentry{RR UB};\label{plot:skr-dir1}
            \addlegendimage{dirls, dir mark*, xi 0.03};\addlegendentry{RR UB};\label{plot:skr-dir3}

            \addplot [rrsls, xi 0.01, no marks] table [x index=0, y index=1] {./data_skr_collective_xi0.01.csv};
            \addplot [rrsls, xi 0.03, no marks] table [x index=0, y index=1] {./data_skr_collective_xi0.03.csv};
            
            \addplot [rrhls, xi 0.01, no marks] table [x index=0, y index=2] {./data_skr_collective_xi0.01.csv};
            \addplot [rrhls, xi 0.03, no marks] table [x index=0, y index=2] {./data_skr_collective_xi0.03.csv};
            
            \addplot [dirls, xi 0.01, no marks] table [x index=0, y index=5] {./data_skr_collective_xi0.01.csv};
            \addplot [dirls, xi 0.03, no marks] table [x index=0, y index=5] {./data_skr_collective_xi0.03.csv};

            \addplot [only marks, xi 0.01, rrs mark, filtered plot] table [x index=0, y index=1] {./data_skr_collective_xi0.01.csv};
            \addplot [only marks, xi 0.03, rrs mark*, filtered plot] table [x index=0, y index=1] {./data_skr_collective_xi0.03.csv};
                                         
            \addplot [only marks, xi 0.01, rrh mark, filtered plot] table [x index=0, y index=2] {./data_skr_collective_xi0.01.csv};
            \addplot [only marks, xi 0.03, rrh mark*, filtered plot] table [x index=0, y index=2] {./data_skr_collective_xi0.03.csv};
                                         
            \addplot [only marks, xi 0.01, dir mark, filtered plot] table [x index=0, y index=5] {./data_skr_collective_xi0.01.csv};
            \addplot [only marks, xi 0.03, dir mark*, filtered plot] table [x index=0, y index=5] {./data_skr_collective_xi0.03.csv};

            \addplot [only marks, xi 0.03, rrh mark*, mark repeat=1] coordinates {
                (44, 1.279482e-03)
                (48, 1.249502e-04)
                (49, 1e-5)
                (49, 1e-6)
            };

            \addplot [only marks, xi 0.03, dir mark*, mark repeat=1] coordinates {
                (98, 3.953949e-05)
                (100, 4e-6)
                (100, 3e-7)
            };

            \addplot [only marks, xi 0.03, rrs mark*, mark repeat=1, mark phase=1] coordinates {
                (95, 9.848922e-05)
                (99, 1.305158e-05)
                (99, 1e-6)
            };

            \addplot [only marks, xi 0.01, rrh mark, mark repeat=1] coordinates {
                (180, 5.264488e-07)
            };
        \end{semilogyaxis}
    \end{tikzpicture}
    \caption{Secret key rate against collective attacks, computed with different excess noise values, namely $\xi=0.01$ (blue lines with empty markers) and $\xi=0.03$ (orange lines with filled markers). We assumed a linear quantum channel model.\label{fig:skr-collective}.}
\end{figure}

\subsection{Bit Error Rate}\label{sec:simulations-results-ber}

\begin{figure*}
    \centering
    \ref{legend:BER-PAM4-half}\\
    \vspace{-.5em}
    \subfloat{
        \begin{tikzpicture}%
            \begin{semilogyaxis}[
                ymax=.5,
                ymin=1e-6,
                width=\columnwidth,
                xlabel={$E_b/N_0$ [dB]},
                ylabel={BER},
                every tick label/.append style={
                    font=\ticklabelsize
                },
                legend to name={legend:BER-PAM4-half},
                legend columns=5,
                every axis legend/.append style={
                    /tikz/every even column/.append style={
                        column sep=.5em
                    }
                },
                grid=major
            ]
                \addplot [black, no marks, dirls, dir colour] table [
                    x={SNR_DIR},
                    y={BER_DIR},
                    col sep=comma
                ] {./data_pam4_ber_r1d2.csv}; \addlegendentry{DR}

                \addplot [black, no marks, rrsls, rrs colour] table [
                    x={SNR_RRS_U},
                    y={BER_RRS_U},
                    col sep=comma
                ] {./data_pam4_ber_r1d2.csv}; \addlegendentry{RRS}

                \addplot [black, no marks, rrhls, rrh colour] table [
                    x={SNR_RRH_U},
                    y={BER_RRH_U},
                    col sep=comma
                ] {./data_pam4_ber_r1d2.csv}; \addlegendentry{RRH}

                \addplot [black, no marks, dashdotdotted, forget plot] table [
                    x=EsN0dB,
                    y=ber,
                    col sep=comma
                ] {./data_ber_c05_alpha0.65.csv}; %

                \node [outer sep=0, inner sep=0, rotate=-85] (citeNode) at (axis cs: 3.6, 1e-4) {RRS \cite{origlia2025soft}};

            \end{semilogyaxis}
            \node [anchor=south west, align=left] at (.1,.1) {(a)\\4-PAM\\$R=1/2$};
        \end{tikzpicture}
        \label{fig:pam4-r1/2}
    }~%
    \subfloat{
        \begin{tikzpicture}%
            \begin{semilogyaxis}[
                ymax=.5,
                ymin=1e-6,
                width=\columnwidth,
                xlabel={$E_b/N_0$ [dB]},
                ylabel={BER},
                every tick label/.append style={
                    font=\ticklabelsize
                },
                grid=major
            ]
                \addplot [black, no marks, dirls, dir colour] table [
                    x expr=\EsToEb{SNR_DIR}{.5},
                    y={BER_DIR},
                    col sep=comma
                ] {./data_pam4_ber_r1d4.csv};

                \addplot [black, no marks, rrsls, rrs colour] table [
                    x expr=\EsToEb{SNR_RRS_U}{.5},
                    y={BER_RRS_U},
                    col sep=comma
                ] {./data_pam4_ber_r1d4.csv};

                \addplot [black, no marks, rrhls, rrh colour] table [
                    x expr=\EsToEb{SNR_RRH_U}{.5},
                    y={BER_RRH_U},
                    col sep=comma
                ] {./data_pam4_ber_r1d4.csv};

            \end{semilogyaxis}
            \node [anchor=south west, align=left] at (.1,.1) {(b)\\4-PAM\\$R=1/4$};
        \end{tikzpicture}
        \label{fig:pam4-r1/4}
    }
    \\[.5em]
    \vspace{-1em}
    \subfloat{
        \begin{tikzpicture}%
            \begin{semilogyaxis}[
                ymax=.5,
                ymin=1e-6,
                width=\columnwidth,
                xlabel={$E_b/N_0$ [dB]},
                ylabel={BER},
                every tick label/.append style={
                    font=\ticklabelsize
                },
                grid=major
            ]
                \addplot [black, no marks, dirls, dir colour] table [
                    x expr=\EsToEb{SNR_DIR}{1.5},
                    y={BER_DIR},
                    col sep=comma
                ] {./data_pam8_ber_r1d2.csv};

                \addplot [black, no marks, rrsls, rrs colour] table [
                    x expr=\EsToEb{SNR_RRS_U}{1.5},
                    y={BER_RRS_U},
                    col sep=comma
                ] {./data_pam8_ber_r1d2.csv};

                \addplot [black, no marks, rrhls, rrh colour] table [
                    x expr=\EsToEb{SNR_RRH_U}{1.5},
                    y={BER_RRH_U},
                    col sep=comma
                ] {./data_pam8_ber_r1d2.csv};

            \end{semilogyaxis}
            \node [anchor=south west, align=left] at (.1,.1) {(c)\\8-PAM\\$R=1/2$};
        \end{tikzpicture}
        \label{fig:pam8-r1/2}
    }~%
    \subfloat{
        \begin{tikzpicture}%
            \begin{semilogyaxis}[
                ymax=.5,
                ymin=1e-6,
                width=\columnwidth,
                xlabel={$E_b/N_0$ [dB]},
                ylabel={BER},
                every tick label/.append style={
                    font=\ticklabelsize
                },
                grid=major
            ]
                \addplot [black, no marks, dirls, dir colour] table [
                    x expr=\EsToEb{SNR_DIR}{.75},
                    y={BER_DIR},
                    col sep=comma
                ] {./data_pam8_ber_r1d4.csv};

                \addplot [black, no marks, rrsls, rrs colour] table [
                    x expr=\EsToEb{SNR_RRS_U}{.75},
                    y={BER_RRS_U},
                    col sep=comma
                ] {./data_pam8_ber_r1d4.csv};

                \addplot [black, no marks, rrhls, rrh colour] table [
                    x expr=\EsToEb{SNR_RRH_U}{.75},
                    y={BER_RRH_U},
                    col sep=comma
                ] {./data_pam8_ber_r1d4.csv};

            \end{semilogyaxis}
            \node [anchor=south west, align=left] at (.1,.1) {(d)\\8-PAM\\$R=1/4$};
        \end{tikzpicture}
        \label{fig:pam8-r1/4}
    }\vspace{-1em}
    \caption{BER plots for 4-PAM and 8-PAM at $R=1/2$ and $R=1/4$. For reference, in each sub-plot we report the BER curve of the DR, as the corresponding SKR (\figref{fig:mutual-information-pam4} for 4-PAM) is an upper bound for the SKR in the RRS setting.}\label{fig:ber}
\end{figure*}
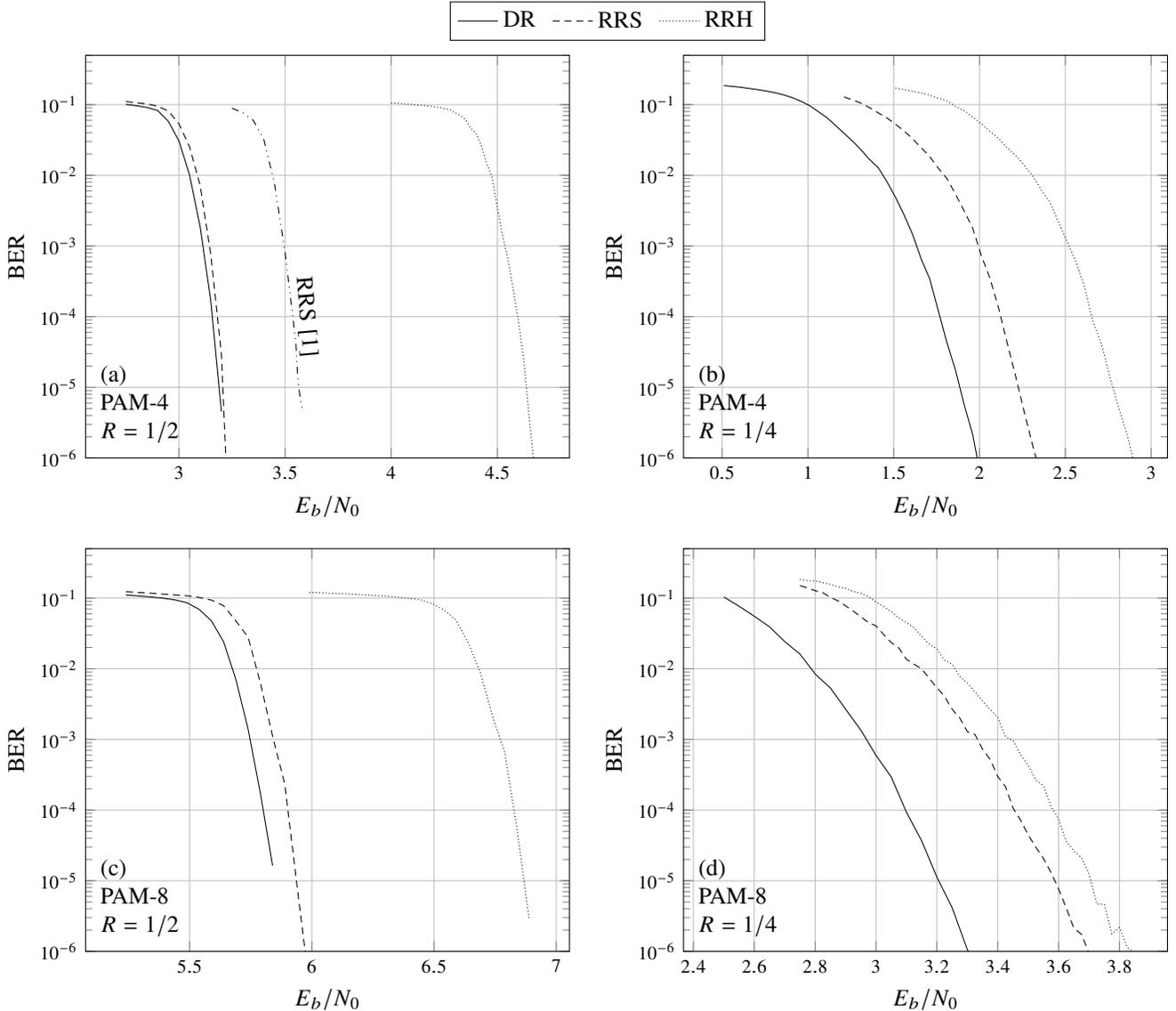

A second set of simulations shows the performance of the proposed RRS scheme in terms of bit error rate (BER) for 4-PAM and 8-PAM coded transmission over an AWGN channel. In these simulations, we consider Gray mapping, DVB-S2 LDPC codes with rates 1/2 and 1/4 and a blocklengh of 64800 \cite{dvbs2_standard}, and implement the RRS scheme with adaptive thresholds and alternating configurations ($C^{[5]}$ for 4-PAM and $C^{[85]}$ for 8-PAM). The LDPC decoder~\cite{origlia_2025_15221950} implements the sum-product algorithm \cite{ryan2009channel, mackay2003information} and takes into account the syndrome when processing check node inputs to produce the messages for the connected variable nodes. Simulation parameters are summarized in Table~\ref{tab:sim-param}.

\begin{table}[!tb]
    \centering
    \caption{Simulation Parameters}\label{tab:sim-param}
    \vspace{-.75em}
    
\begin{tabular}{|r|l|}
    \hline
    %% \multicolumn{2}{|c|}{TABLE~\ref{tab:sim-param}}\\\hline
    Modulations & 4-PAM, 8-PAM \\\hline
    Code rates & 1/2, 1/4\\\hline
    Block length & 64800\\\hline
    Max LDPC iterations & 50\\\hline
    % Number of frames & \numrange{250}{5000}\\\hline
    % Min. number of frame errors & 250\\\hline
\end{tabular}

\end{table}

The corresponding BER curves are shown in \figref{fig:ber}.
As benchmarks, we also report the BER obtained with the RRH scheme and with direct reconciliation, where Bob recovers Alice's sequence using the channel output $Y$ and the syndrome, as in a classical communication scheme. We take the latter as a reference because it represents the case where soft information is fully available at the decoder, and the corresponding SKR is equal to the upper bound for the SKR of RRS.

With a \mbox{rate-1/2} code employed in the 4-PAM scenario (\figref{fig:pam4-r1/2}), RRH has a gap of about \num{1.4}\unit{\dB} from DR. This large gap is almost completely filled by RRS, which lies within \num{0.04}\unit{\dB} from DR.
This is consistent with the behavior observed for the SKR curves in \figref{fig:mutual-information-pam4-zoom}.
For reference, in \figref{fig:pam4-r1/2} we also report the result we obtained in our recent work \cite{origlia2025soft}, where we a gain of roughly 1dB w.r.t. RRH was observed, further improved by resolving numerical issues.

With a \mbox{rate-1/4} code (\figref{fig:pam4-r1/4}), the gap between RRH and DR reduces to about \num{0.9}\unit{\dB}, and is only partially filled by RRS, which however still provides a relevant gain of more than \num{0.5}\unit{\dB} over RRH.
The gains of the RRS scheme over RRH and its residual gap to DR are summarized in table \ref{tab:pam-4-r1d4-gap-gain}.  For reference, gaps and gains measured on the SKR curves in \figref{fig:mutual-information-pam4-zoom} are reported, too. The gains are roughly the same, but the residual gaps appear  larger in the BER curves than in the SKR curves, particularly at lower rate.
This larger residual gap may be caused by the mismatched bit-wise decoding implemented by the belief propagation decoder, and could potentially be reduced through symbol-wise decoding, non-binary codes, or optimized bit-labeling strategies. This aspect is currently under investigation and will be addressed in future work.

\begin{table}[!tb]
    \centering
    \caption{Gap Comparison}\label{tab:pam-4-r1d4-gap-gain}
    \vspace{-.75em}
    
\begin{tabular}{|rl|c|c|c|}
    \hline
    \multicolumn{2}{|l|}{%TABLE~\ref{tab:pam-4-r1d4-gap-gain}
    }%
    % &
    & Gain [dB] & Residual gap [dB] \\\hline\hline
    % \multirow{2}{2em}{(F)}& BER & 0.59 & 0.46\\
    % & MI & 0.6 & 0.09\\\hline
    BER & ($10^{-3}$, $R=1/2$) & 1.39 & 0.04 \\
    SKR & ($1$ b/cu)           & 1.2 & 0 \\\hline
    BER & ($10^{-3}$, $R=1/4$) & 0.53 & 0.36\\
    SKR & ($0.5$ b/cu)         & 0.63 & 0.07\\\hline
\end{tabular}

\end{table}

Regarding the 8-PAM modulation, at rate $R=1/2$ (\figref{fig:pam8-r1/2}) the BER of our scheme lays within \num{0.1}\unit{\dB} from the BER of the DR scheme.
When a \mbox{rate-1/4} code is employed (\figref{fig:pam8-r1/4}), the improvement is less strong, but the RRS still has a positive impact.
In fact, despite the fact that the gap to the BER of the direct reconciliation amounts to about \num{0.35}\unit{\dB}, a gain of at least \num{0.1}\unit{\dB} with respect to RRH is still present.

\section{Conclusions}\label{sec:conclusions}
In this work we introduced a novel technique that improves the efficiency of RR for CV-QKD protocols that use PAM or QAM alphabets.
In this approach, denoted as RRS, Bob computes and discloses a carefully designed soft metric that assists Alice in reconciling her transmitted symbols with Bob's decisions (the key), while not revealing any information about such decisions to a potential eavesdropper.
    We formalized the concept of \emph{configurations} underlying the construction of the soft metric and  analyzed how the choice of the configurations, along with the thresholds used to derive the key from Bob's measurements, %
    affect the achievable SKR.%

    We demonstrated that the RRS scheme exhibits a significant gain over RR without the soft metric (the RRH scheme) %
    in terms of the achievable SKR, both when no eavesdropper is present on the quantum channel, %
    and when it performs collective attacks in the asymptotic regime on a linear quantum channel. %

Furthermore, we implemented and tested a complete version of the proposed RRS scheme, employing binary LDPC codes and bit-wise belief-propagation decoding.
Although the BER results generally follow the SKR trends, the observed gains and gaps do not totally reflect those in the SKR%
. The reason for this discrepancy is currently under investigation, but a likely explanation is the use of bit-wise decoding%
    , whose performance is better predicted by the normalized generalized mutual information (GMI) %
    \cite{alvarado_replacing_2015}%
.

In future work, we will compute the GMI of RRS to better understand the impact of symbol-to-bit mapping and bit-wise decoding on performance, and possibly to close the gap between hard and soft decoding also in terms of BER. We will also explore probabilistic constellation shaping and threshold optimization, and focus on the low-rate regime of particular interest for QKD. 

\section*{Acknowledgement}
The authors sincerely thank Erdem Eray Cil\orcidlink{0000-0003-3241-4446} (Karlsruhe Institute of Technology) for useful discussions about security proofs.

\appendix

\subsection{Equivalent Configurations and Transformation Functions}\label{sec:appendix-configurations}
In the following we will refer to the soft metric produced by the scheme employing a configuration $C^{[b]}$ with $N^{[b]}$.

\subsubsection{Flipped functions}
A pair $C^{[b]}$ and $C^{[b^{(F)}]}$ of flipped configurations has opposite monotonicity directions in each decision interval, i.e.,
\begin{equation}\label{eq:config-def-flipped}
 \forall i \in \{1, \ldots, M\} : b_i^{(F)} = \bar{b_i} \Rightarrow G_i^{[b^{(F)}]} = 1-G_i^{[b]}.   
\end{equation}
For instance, in the 4-PAM scenario, configurations $C^{[2]}$ and $C^{[13]}$ are equivalent by flipping, and the corresponding transformation functions would look like this:
\begin{align}
    G^{[4]} &\sim [\shortarrow{1},\shortarrow{1},\shortarrow{-1},\shortarrow{1}]\\
    G^{[11]} &\sim [\shortarrow{-1},\shortarrow{-1},\shortarrow{1},\shortarrow{-1}]
\end{align}

\begin{IEEEproof}{(Equivalence of flipped functions)}
    From \eqref{eq:config-def-flipped} it follows that \mbox{${G_i^{[b^{(F)}]}}^{-1}(n^{[b^{(F)}]}) = {G_i^{[b]}}^{-1}(n^{[b]})$} and for any $y\in D_i$: \mbox{$|{G_i^{[b^{(F)}]}}'(y)| = |{G_i^{[b]}}'(y)|$}. This implies that for the flipped configuration it holds that:
    \begin{equation}
        \begin{split}
            f_{N^{[b^{(F)}]},\Xhat|X}(n^{[b^{(F)}]}, a_i|a_j) = \frac{
            f_{Y|X}({G_i^{[b^{(F)}]}}^{-1}(n^{[b^{(F)}]})|a_j)
        }{
            |{G_i^{[b^{(F)}]}}'({G_i^{[b^{(F)}]}}^{-1}(n^{[b^{(F)}]})|
        } =\\ \frac{
            f_{Y|X}({G_i^{[b]}}^{-1}(n^{[b]})|a_j)
        }{
            |{G_i^{[b]}}'({G_i^{[b]}}^{-1}(n^{[b]})|
        } = f_{N^{[b]},\Xhat|X}(n^{[b]}, a_i|a_j),
        \end{split}
    \end{equation}
    and the distributions above being equal, $H(\Xhat|X, N)$ does not change when we replace $N^{[b^{(F)}]}$ by $N^{[b]}$ in \eqref{eq:cond-entropy}. Moreover, $H(\Xhat)$ in \eqref{eq:mi-expanded} does not depend on the configuration. So the mutual information of RRS that employs either of them does not change, i.e.,
    \begin{equation}\label{eq:mi-identity-flipped}
        I(\Xhat;X | N^{[b^{(F)}]}) = I(\Xhat;X | N^{[b]}).
    \end{equation}
\end{IEEEproof}

\subsubsection{Mirrored functions}
A mirrored function is obtained by mirroring a transformation function, i.e., 
\begin{equation}\label{eq:config-def-mirrored}
    G_i^{[b^{(M)}]}(y) = G_{M+1-i}^{[b]}(-y).
\end{equation}
For instance, in the 4-PAM scenario, configurations $C^{[4]}$ and $C^{[13]}$ are equivalent by mirroring, as the corresponding transformation functions would look like this:
\begin{align}
    G^{[4]} &\sim [\shortarrow{1},\shortarrow{1},\shortarrow{-1},\shortarrow{1}]\\
    G^{[13]} &\sim [\shortarrow{-1},\shortarrow{1},\shortarrow{-1},\shortarrow{-1}]
\end{align}

In the following we will assume that the input distribution and the decision regions $D_i$ are symmetric, i.e.,
\begin{align}
    \forall a_i \in\Abet: \; \PP{X=a_i} = \PP{X = -a_i = a_{M+1-i}}\\
    \inf D_i = \sup D_{M+1-i}, \quad \sup D_i = \inf D_{M+1-i},
\end{align}
so $y\in D_i \Longleftrightarrow -y \in D_{M+1-i}$.

\begin{IEEEproof}{(Equivalence of mirrored functions)}
    The first step is to prove that
    \begin{equation}\label{eq:config-mirrored-distributions}
        f_{N^{[b^{(M)}]},\Xhat|X}(n, a_i | a_j) = f_{N^{[b]},\Xhat|X}(n, a_{M+1-i} | a_{M+1-j}).
    \end{equation}
    
    We can immediately establish the following relationships from definition \eqref{eq:config-def-mirrored}
    \begin{align}
        {G_i^{[b^{(M)}]}}^{-1}(n) &= -{G_{M+1-i}^{[b]}}^{-1}(n)\label{eq:G3_inverse},\\
        {G_i^{[b^{(M)}]}}'(y) &= -{G_{M+1-i}^{[b]}}'(-y).\label{eq:G3_derivative}
    \end{align}
    By applying \eqref{eq:G3_derivative} to \eqref{eq:G3_inverse}, we obtain
    \begin{equation}\label{eq:G3_abs_nesting}
        |{G_i^{[b^{(M)}]}}'({G_i^{[b^{(M)}]}}^{-1}(n))| = |{G_{M+1-i}^{[b]}}'({G_{M+1-i}^{[b]}}^{-1}(n))|
    \end{equation}

    By using the symmetry of the PAM constellation, i.e., $-a_j = a_{M+1-j}$, and \eqref{eq:G3_inverse} again, we obtain
    \begin{align}
        \left({G_i^{[b^{(M)}]}}^{-1}(n) - a_j\right)^2 = \left({G_{M+1-i}^{[b]}}^{-1}(n) - a_{M+1-j}\right)^2\\
        \Rightarrow f_{Y|X}({G_i^{[b^{(M)}]}}^{-1}(n) | a_j) = f_{Y|X}({G_{M+1-i}^{[b]}}^{-1}(n) | a_{M+1-j}).\label{eq:conditional-gaussian-cross}
    \end{align}
    Dividing \eqref{eq:conditional-gaussian-cross} by \eqref{eq:G3_abs_nesting} term by term, as in \eqref{eq:n-xhat-cond-x}, we prove \eqref{eq:config-mirrored-distributions}.
    For the second part of the proof, we write the second term of \eqref{eq:mi-expanded} as 
    \begin{equation}\label{eq:mi-mirrored-n3}
        \begin{split}
            H(&\Xhat|X, N^{[b^{(M)}]}) = \int_0^1\sum_{i=1}^{M/2} \sum_{j=1}^{M/2} \bigg(\mathcal{F}_{i,j}^{[b^{(M)}]}(n) +
            \mathcal{F}_{i,M+1-j}^{[b^{(M)}]}(n) + \\ & + \mathcal{F}_{M+1-i,j}^{[b^{(M)}]}(n)+
            \mathcal{F}_{M+1-i,M+1-j}^{[b^{(M)}]}(n)
            \bigg)\text{d}n
        \end{split}
    \end{equation}
    where
    \begin{equation}
        \begin{split}
            \mathcal{F}_{i,j}(n) &= \PPP[X]{a_j}f_{N,\Xhat|X}(n, a_i|a_j) \\
            &\cdot\log_2\matchparen{
                \frac{
                    \sum_{a_k\in\Abet}f_{N,\Xhat|X}(n, a_k|a_j)
                }{
                    f_{N,\Xhat|X}(n, a_i|a_j)
                }
            }.
        \end{split}
    \end{equation}
    Because of \eqref{eq:config-mirrored-distributions} and the symmetries discussed above, we have that
    \begin{equation}
        \mathcal{F}_{i,j}^{[b^{(M)}]}(n) = \mathcal{F}_{M+1-i,M+1-j}^{[b]}(n),
    \end{equation}
    and we simply obtain a re-arrangement of the sum inside the integral in \eqref{eq:mi-mirrored-n3}.
    So, we can conclude that $H(\Xhat|X, N^{[b^{(M)}]}) = H(\Xhat|X, N^{[b]})$, therefore
    \begin{equation}\label{eq:mi-identity-mirrored}
        I(\Xhat ; X | N^{[b^{(M)}]}) = I(\Xhat ; X | N^{[b]})
    \end{equation}
\end{IEEEproof}

We can immediately establish the equivalence of  \eqref{eq:config-def-mirrored} to the following equation
\begin{equation}
    b_i^{(M)} = \bar{b}_{M+1-i}
\end{equation}
as it will be useful for proving the last configuration equivalence.

\begin{remark}
    In Section~\ref{sec:theory} we claimed that the analysis of the scheme holds for any input PMF $\PPP[X]{a_i}$, 
    however in this proof we assumed that opposite constellation points have the same probability.
    This is not a very restrictive assumption, because constellation points probabilities are usually shaped to be symmetric. %
\end{remark}
\begin{remark}\label{remark:mirrored-additive-noise} We proved~\eqref{eq:config-mirrored-distributions} for the Gaussian noise distribution, however the proof holds
for any other additive noise process whose distribution satisfies \eqref{eq:config-mirrored-distributions}.
If this condition is not satisfied, one should simply consider $C^{[b]}$ and $C^{[b^{(M)}]}$ as configurations with distinct performance.
\end{remark}

\subsubsection{Reversed configurations}
In this configuration we just reverse the order of the monotonicities, i.e.,
\begin{equation}\label{eq:config-def-reversed}
    b_i^{(R)} = b_{M+1-i}
\end{equation}

Transformations functions with equivalent monotonicities would look like:
\begin{align}
    G^{[4]} &\sim [\shortarrow{1},\shortarrow{1},\shortarrow{-1},\shortarrow{1}]\\
    G^{[2]} &\sim [\shortarrow{1},\shortarrow{-1},\shortarrow{1},\shortarrow{1}]
\end{align}

\begin{IEEEproof}{(Equivalence of reversed configurations)}

    \begin{align}
        b_i^{(R)} &= b_{M+1-i} =\\
        &=\bar{\bar{b}}_{M+1-i} =\\
        &=\bar{b}_i^{(M)}=\\
        &= (b^{(M)})^{(F)}_i
    \end{align}
    So, to obtain the reversed configuration corresponding to $C^{[b]}$, instead of reversing the order of the direction of the monotonicities, we can equivalently mirror $C^{[b]}$, and then flip the result. We can now apply the identities we found for the corresponding mutual informations.

\begin{align}
     I(\Xhat;X | N^{[b^{(R)}]}) :&= I(\Xhat;X | N^{[(b^{(M)})^{(F)}]}) \\
     &= I(\Xhat;X | N^{[(b^{(M)})]}) \label{eq:mi-identity-flipped-mirrored}\\
     &= I(\Xhat;X | N^{[b]}),\label{eq:mi-identity-mirrored-1}
\end{align}

where in \eqref{eq:mi-identity-flipped-mirrored} we used \eqref{eq:mi-identity-flipped} with configurations $C^{[(b^{(M)})^{(F)}]}$ and $C^{[(b^{(M)})]}$, and in \eqref{eq:mi-identity-mirrored-1} we used \eqref{eq:mi-identity-mirrored} with configurations $C^{[(b^{(M)})]}$ and $C^{[b]}$.
\end{IEEEproof}

\begin{remark}
    The proof of equivalence of reversed configurations is based on the equivalence of mirrored configurations.
    Consequently, the claims of \emph{Remark}~\ref{remark:mirrored-additive-noise}
    apply also to reversed configurations.
\end{remark}

\subsection{List of Equivalent Configurations}\label{appendix:table-equivalent-configs}
In Tables~\ref{tab:equivalent-configurations-bpsk} and \ref{tab:equivalent-configurations} we list the equivalent configurations for BPSK and 4-PAM, respectively, the first element of each row being the "reference" configuration.

\begin{table}[!t]
    \centering
    \caption{Equivalent configurations for BPSK}
    \label{tab:equivalent-configurations-bpsk}
    \vspace{-.75em}

\begin{tabular}{c|lll}
    \hline
    $C^{[b]}$ & $C^{[b^{(F)}]}$ & $C^{[b^{(M)}]}$ & $C^{[b^{(R)}]}$ \\
    \hline
    $C^{[0]}$ & $C^{[3]}$ & $C^{[3]}$ & $C^{[0]}$\\
    $C^{[1]}$ & $C^{[2]}$ & $C^{[1]}$ & $C^{[2]}$\\
    \hline
\end{tabular}

\end{table}

\begin{table}[!t]
    \centering
    \caption{Equivalent configurations for 4-PAM}
    \label{tab:equivalent-configurations}
    \vspace{-.75em}

\begin{tabular}{c|lll}
    \hline
    $C^{[b]}$ & $C^{[b^{(F)}]}$ & $C^{[b^{(M)}]}$ & $C^{[b^{(R)}]}$ \\
    \hline
    $C^{[0]}$ & $C^{[15]}$ & $C^{[15]}$ & $C^{[0]}$\\
    $C^{[1]}$ & $C^{[14]}$ & $C^{[7]}$ & $C^{[8]}$\\
    $C^{[2]}$ & $C^{[13]}$ & $C^{[11]}$ & $C^{[4]}$\\
    $C^{[3]}$ & $C^{[12]}$ & $C^{[3]}$ & $C^{[12]}$\\
    $C^{[5]}$ & $C^{[10]}$ & $C^{[5]}$ & $C^{[10]}$\\
    $C^{[6]}$ & $C^{[9]}$ & $C^{[9]}$ & $C^{[6]}$\\
    \hline
\end{tabular}

\end{table}

Configurations $C^{[2]}$ and $C^{[3]}$ for BPSK, and $C^{[4]}$ and $C^{[7]}, \ldots, C^{[15]}$ for 4-PAM are not listed as reference configurations, because they are equivalent to some other one already listed.
In fact, it is possible to use the fact that flipping, mirroring and order reversal are self-inverse operations.
\begin{IEEEproof}{(Self-inverse property)}
    \begin{align}
        \matchparen{b_i^{(F)}}^{(F)} &= \bar{b}_i^{(F)} = \bar{\bar{b}}_i =b_i\\
        \matchparen{b_i^{(M)}}^{(M)} &= \bar{b}_{M-i+1}^{(M)} = \bar{\bar{b}}_{M - (M-i+1) + 1} = b_i \\
        \matchparen{b_i^{(R)}}^{(R)} &= b_{M-i+1}^{(R)} = b_{M - (M-i+1) + 1} = b_i
    \end{align}
\end{IEEEproof}

In Table~\ref{tab:order-classes} we report the number of distinct equivalent classes for typical modulation orders.
\begin{table}[!t]%
    \centering
    \caption{Number of configuration classes vs. PAM modulation order.}
    \label{tab:order-classes}
    \vspace{-.5em}
    
\begin{tabular}{lr}
        \hline
        PAM Order        & Number of classes\\\hline
        2-PAM (BPSK) & 2 \\
        4-PAM        & 6 \\
        8-PAM        & 72\\
        16-PAM       & 16512\\
        \hline
\end{tabular}

\end{table}

\bibliographystyle{IEEEtran}
\bibliography{IEEEabrv,references}

\end{document}